\documentclass[pdflatex,sn-nature]{sn-jnl}

\usepackage{siunitx}
\usepackage{graphicx}%
\usepackage{multirow}%
\usepackage{amsmath,amssymb,amsfonts}%

\begin{document}

\title [Article Title]{Variable dose slicer for refractive index engineering in two-photon polymerization}

\author*[1]{\fnm{Micha\l{}} \sur{Ziemczonok}} \email{michal.ziemczonok@pw.edu.pl}
\author[2,3]{\fnm{Koen} \sur{Vanmol}}
 
\affil[1]{\orgname{Warsaw University of Technology, Institute of Micromechanics and Photonics}, \orgaddress{\street{Boboli 8 Street}, \city{Warsaw}, \postcode{02-525}, \country{Poland}}}
\affil[2]{\orgname{Brussels Photonics (B-PHOT), Vrije Universiteit Brussel}, \orgaddress{\street{Pleinlaan 2}, \city{Brussel}, \postcode{1050}, \country{Belgium}}}
\affil[3]{\orgname{Flanders Make@VUB - BP\&M}, \orgaddress{\street{Pleinlaan 2}, \city{Brussel}, \postcode{1050}, \country{Belgium}}}

\abstract{In two-photon polymerization (TPP), the degree of conversion (DC) of the resin has an effect on a broad range of material properties like refractive index (RI) or stiffness. Heterogeneous DC can substitute material doping and multimaterial structures, and outright enable structure designs not possible otherwise. However, obtaining variable DC in the polymer, typically achieved by implementing a variable exposure dose, is held back due to the lack of software support for fabrication and measurement techniques for validation, adding up to a high barrier of entry.

This work presents two major breakthroughs in (3+1)D TPP: design freedom and variable-DC fabrication, that are provided by an open-source slicer, as well as calibration methodology for determination of the RI for any DC. Application examples include grayscale lithography, control over writing direction and trajectories, as well as bio-mimicking microphantoms with carefully engineered 3D RI. 
Results of the RI calibration demonstrate excellent repeatability, accuracy and stability of variable-DC structures. Supported by in-depth metrological analysis, the goal is to popularize variable DC printing within TPP community and to get more out of the existing TPP systems and workflows. In summary, this work provides a complete toolbox for 3D RI engineering and sets the stage for new inventions enabled by point-wise dose control.}


\maketitle
\section{Introduction}
Two-photon polymerization (TPP) is a microfabrication technique revolutionizing research and industry in microoptics, integrated photonics, micromechanics and life sciences \cite{Harinarayana2021,OHalloran2022,Wang2024}. 
Also called multiphoton/laser lithography or direct laser writing, it relies on non-linear light absorption to induce polymerization of the photosensitive material along the focused beam path. 
High-NA optics and ultrashort laser pulses are leveraged to keep the interaction volume small in both lateral and axial directions, enabling fabrication of submicrometer three-dimensional features within millimeter or even centimeter scale structures.
Unmatched synergy of fabrication speed \cite{Hahn2020}, geometrical accuracy \cite{LaFratta2017,Vora2020} and structure functionalization through material science and resin composition \cite{Carlotti2019,Liao2020,Yang2021,OHalloran2022,Jaiswal2023,Song2024,ArriagaDvila2025} makes TPP extremely attractive in a multitude of applications. 

One of the crucial process parameters in TPP is the interplay between the energy dose and the resulting degree of conversion (DC) of the resin. In fact the first print that is typically executed after any major change in the TPP process is called a dose test, that aims to find a suitable processing window of scan speed and laser power, where at low accumulated dose there is barely any cross-linking of the polymer and emerged structures lack structural integrity, while too high dose leads to microexplosions and bubble formation. The distance between subsequent lines and layers, so-called hatching and slicing, is also carefully adjusted depending on the TPP voxel size \cite{Guney2016,Bougdid2020,Drobecq2025}. 
So, in general, system configuration and process parameters are optimized to maximize fabrication speed or shape accuracy, whereas material science and resin composition are typically responsible for functionalization of the structures. Many applications and needs can be addressed simply by using different geometries of the structures and their base material's properties. 
However, many researchers noticed that the DC offers an additional degree of freedom and coined the term (3+1)D printing, since DC is related to a broad range of material properties which opens up new opportunities. 
For example, resins yield greater density in the polymer phase than in the monomer phase, so the DC is also proportional to the elastic modulus \cite{Eren2025} and refractive index (RI) \cite{Zukauskas2015}, and the chemical reactions following two-photon absorption are reflected in Raman spectra \cite{Jiang2014} and fluorescent photobleaching \cite{Sharaf2023}. 
Therefore, this approach has been utilized in stimulus-responsive structures \cite{Huang2020} including light-induced flow to propel microrobots \cite{Konara2025}, photothermal actuators \cite{Hippler2019,Yarali2025,Chen2025,Zhu2025} and bi-materials \cite{Salah2025}, e.g. by crafting microstructures with a negative thermal expansion coefficient \cite{Qu2017} that exploit the difference in this coefficient due to the DC and optimized bi-material lattice. 
Another avenue is control over swelling and mechanical stability \cite{VanHoorick2017,Ugarak2022}, grayscale lithography that dynamically adjusts voxel size to better fit the intended geometry and minimize staircase effects \cite{Aderneuer2021,Khonina2024}, as well as local ablation \cite{Ye2025}. 
RI engineering already reinforces the strong position of TPP for the fabrication of microoptics \cite{Ristok2020,GonzalezHernandez2021,Wang2024,Liu2025}. It enables the fabrication of gradient-index (GRIN) optics, waveguides and volumetric holograms \cite{Zukauskas2015, GonzalezHernandez2023, Porte2021} essential in photonic integration, miniaturized optical systems placed directly on the optical fibers and advanced wavefront shaping. Finally, variable DC also finds applications in imaging system metrology, where the RI distribution is used to create phantoms that resemble biological cells and their organelles \cite{Micha2019}, adjust light scattering strength that obfuscates imaging targets \cite{Lamont2020,Krauze2022,Kim2025} or even mimic 3D RI distributions of organoids \cite{Ziemczonok2025}. 

There are three distinct reasons why the advantages of DC are well-known but rarely utilized: (1) the lack of a standardized file format, as standard file formats for 3D-printing like .stl and .stp only support geometric and topological information with no means of encoding point-wise properties; (2) the calibration of process parameters and resulting DC requires specialized methodologies and measurement equipment; and (3) low-DC regions are mechanically unstable and photosensitive, which limits their reliability and requires careful handling. Nevertheless, in many cases adjustable DC can be an alternative to material doping and it can eliminate the need for multiple photoresins and writing sequences, while in some cases it is preferred or even the only way to bring an idea to life.

This work addresses point (1) by providing an open-source slicer for voxelized structure design tailored for TPP along with a range of practical features and application examples, as well as point (2) with a calibration methodology for the determination of the RI for any DC with extremely high accuracy and precision. These tools and methods aim to popularize variable DC printing and enable researchers and users to fully take advantage of this additional degree of freedom of their TPP systems. Various application avenues highlighted in this paper only scratch the surface of vast design possibilities enabled by precise dose control, especially when tied with mature hardware platforms and fully supported in software by the industry.

\section{Materials and methods}
\subsection{Two-photon polymerization fabrication}
All microstructures were fabricated via TPP lithography using the Photonic Professional GT2 system (Nanoscribe GmbH \& Co. KG, Karlsruhe, Germany), equipped with a 1.4~NA 63x microscope objective, a piezo stage for vertical positioning of the sample, and a dual-axis galvo for lateral scanning. Unless stated otherwise, the structures were fabricated on top of a silanized \qtyproduct[product-units = single]{22 x 22 x 0.17}{\milli\meter} high precision coverslip using the IP-Dip2 resin (Nanoscribe GmbH \& Co. KG, Karlsruhe, Germany). Hatching and slicing distances were set to 0.15 and \SI{0.3}{\micro \meter} respectively, the scan speed was \SI{10000}{\micro \meter \per \second} and the printing was performed in the dip-in configuration. The DC has been modulated via the laser power of the TPP system, given as a percentage (\%) of the \SI{50}{\milli \watt} mean optical power entering the aperture of the objective. After fabrication, the samples were developed in PGMEA (Propylene glycol monomethyl ether acetate; 12 min; Sigma Cat: 484431) followed by Novec 7100 Engineered Fluid (1 min; Sigma Cat: SHH0002) and then air-dried.

\subsection{Refractive index measurements and laser power calibration} \label{sec:methodologyRIcalibration}
The refractive index of the microstructures is determined using digital holographic microscopy (DHM) \cite{Kim2011} in transmission configuration. DHM is a quantitative phase imaging technique with proven interferometric accuracy \cite{Besaga2019,Emery2021,Chaumet2024}. The measured phase shift of light introduced by the sample ($\Delta\varphi$, Eq.~\ref{eq:phase}) is related to its thickness and the RI ($\Delta RI$ - refractive index contrast, $\Delta d$ - height difference, $\lambda$ - wavelength of light). 
\begin{equation}
\Delta\varphi = \Delta RI \Delta d \frac{2\pi}{\lambda}
\label{eq:phase}
\end{equation}
Therefore, the phase and height measurements of the sample in a known RI medium enable calculation of the resin RI. However, performing accurate height measurements of the structures containing light-sensitive resin in a non-destructive manner is challenging. An alternative approach is to use two DHM measurements of the sample mounted in two different RI media, so the RI of the polymer is calculated from Eq.~\ref{eq:phase} using only phase and medium RI values as:
\begin{equation}
RI_{resin} = \frac{\Delta\varphi_1 RI_{immersion2}-\Delta\varphi_2 RI_{immersion1}}{\Delta\varphi_1-\Delta\varphi_2},
\label{eq:rimatching}
\end{equation}
assuming that the wavelength, sample geometry and measurement conditions remain unchanged. More details about the accuracy, repeatability and uncertainty of the RI can be found in Section~\ref{sec:error}. 

The measurement system is based on an off-axis Mach-Zehnder interferometer equipped with a 633~nm laser diode, a 1.3~NA oil immersion objective and a tube lens imaging the sample plane on the camera sensor. Details regarding DHM system configuration can be found in \cite{Desissaire2025}. Phase retrieval has been performed numerically by filtering the +1 diffraction order in the Fourier space and aberration-corrected by subtracting the phase distribution without any objects in the field of view. 50 holograms were captured for each measurement in order to average the phase maps and reduce noise. 

\begin{figure}[h]
    \includegraphics[width=.48\textwidth]{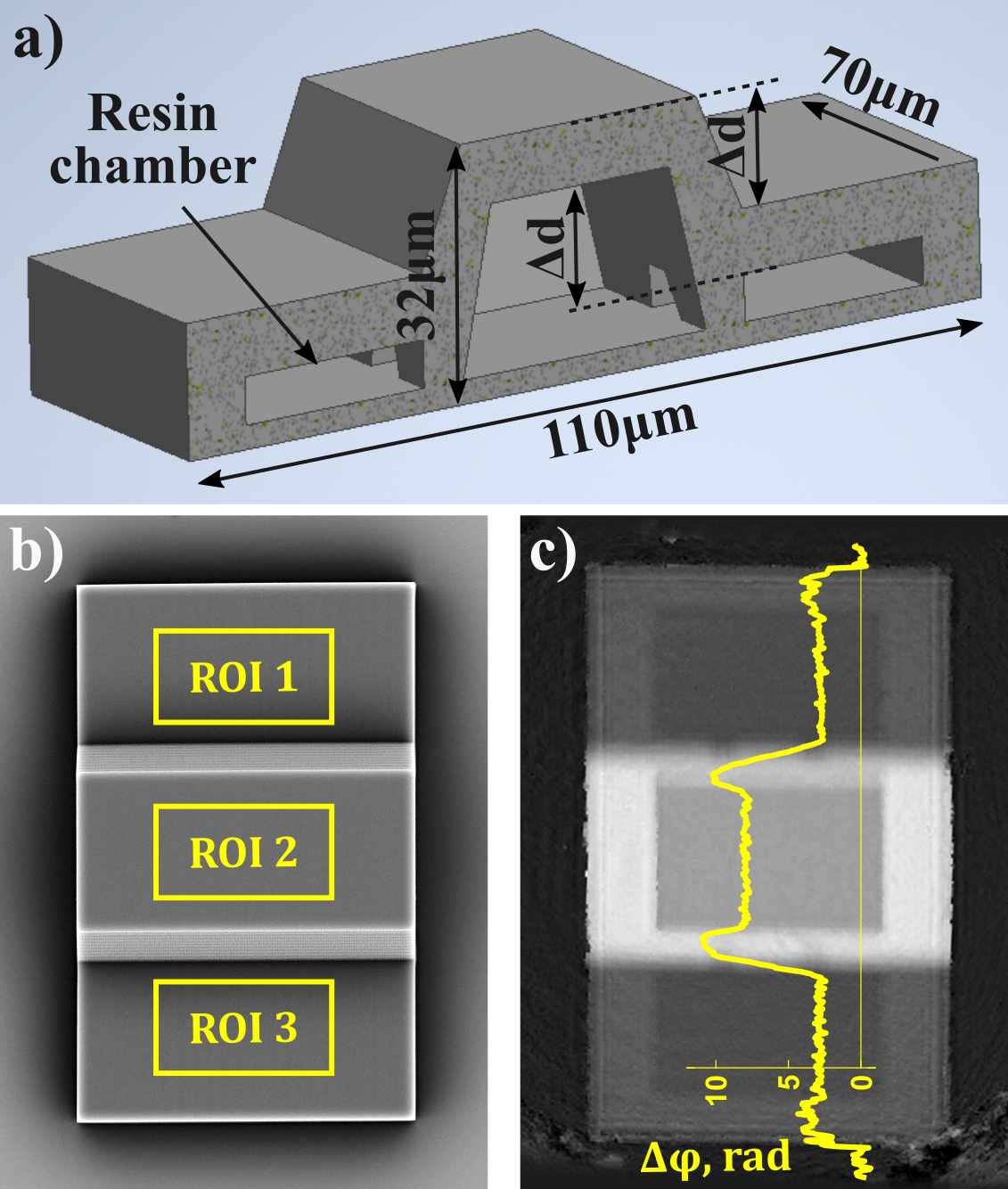} 
    \caption{Structure design for RI calibration of the material. a) Half-section view of the step structure, where the shell printed with high DC protects the resin chamber, that can be polymerized to any degree. The proposed geometry introduces $\Delta$d = \SI{15}{\micro \meter} height difference between the resin and surrounding medium. b) Scanning electron microscopy image of the step with marked regions of interest (ROI) used for analysis. c) Phase map and cross-section of the step structure filled with liquid resin, immersed in 1.50 RI medium and measured by DHM.}
    \label{fig:riramps}
\end{figure}
The calibration structure design is presented in Fig.~\ref{fig:riramps}. It is a three-step structure, where the protective shell is printed with high DC and the internal compartment can be left unprocessed or polymerized to any degree. This geometry facilitates optimal differential phase measurements, that is averaging the phase value in three regions of interest (ROI, Fig.~\ref{fig:riramps}b) and then calculating the difference as $\Delta\varphi = ROI_2-(ROI_1+ROI_3)/2$. 
Provided that these three regions are of equal size and symmetrical relative to the structure outline, the $\Delta\varphi$ calculation compensates for any offset or tilt that is often encountered in phase images. Additionally, small tunnels connecting the chambers are designed to equalize pressure, so with all other geometries being equal, the only phase delay difference between the ROIs originates from the $\Delta d$ portion of the chamber. Therefore the proposed target design, together with the double-immersion approach that eliminates the need to measure the geometry of the sample, makes the whole RI calibration methodology quite accurate and robust. 

Calibration results utilizing a DHM, two immersion mediums and an array of such structures, each printed with different power within the chamber region, are presented in Section~\ref{sec:RIcalibration}. The step structure design is uploaded to the data repository -- see Data Availability section.

\subsection{MATLAB slicer with dynamic printing power adjustments}
The slicer enabling fabrication of arbitrary 3D DC distributions (and thus RI) within the volume of the sample is open-source (see Data Availability section). The script has been written in MATLAB (R2018b). Various code snippets and examples provided in the repository demonstrate basic use and features, some of which are also further described in the Results section. Typical TPP design file formats like grayscale images or .STL files can also be parsed using this slicer after voxelization. In case of images the voxelization is straightforward and an example is included in the repository, while the .STL files need to be converted to the 3D binary mask using e.g. community tools \cite{stl_FEX}. 

In general, the script takes a 3D array of RI values as an input, and outputs a .gwl file, which is compatible with Nanoscribe's processing software DeScribe and their TPP systems. The output format is a list of 4D coordinates (x, y, z, power) which represent the 3D position to be addressed by the scanners followed by the assigned laser power value. Output data format can easily be adjusted to accommodate other manufacturers or custom systems. 
Additional control parameters required for slicing include sampling (y, x, z) of the input 3D RI, slicing distance, hatching distance and a calibration file defining the relationship between the RI and the TPP laser power. Optional parameters include printing direction (enabling parts of the structure to be printed with perpendicular scanning) or structure rotation, further described in the Results section. Moreover, with simple code modifications or \textit{find and replace} function in the text editor it is possible to generate .gwl files with parametric values, enabling convenient parameter sweeps in selected regions of the structure, just like in the case of the RI calibration step in Fig.~\ref{fig:riramps}. 

\section{Results}
\subsection{Refractive index calibration} \label{sec:RIcalibration}
\begin{figure}[htbp]
    \includegraphics[width=.48\textwidth]{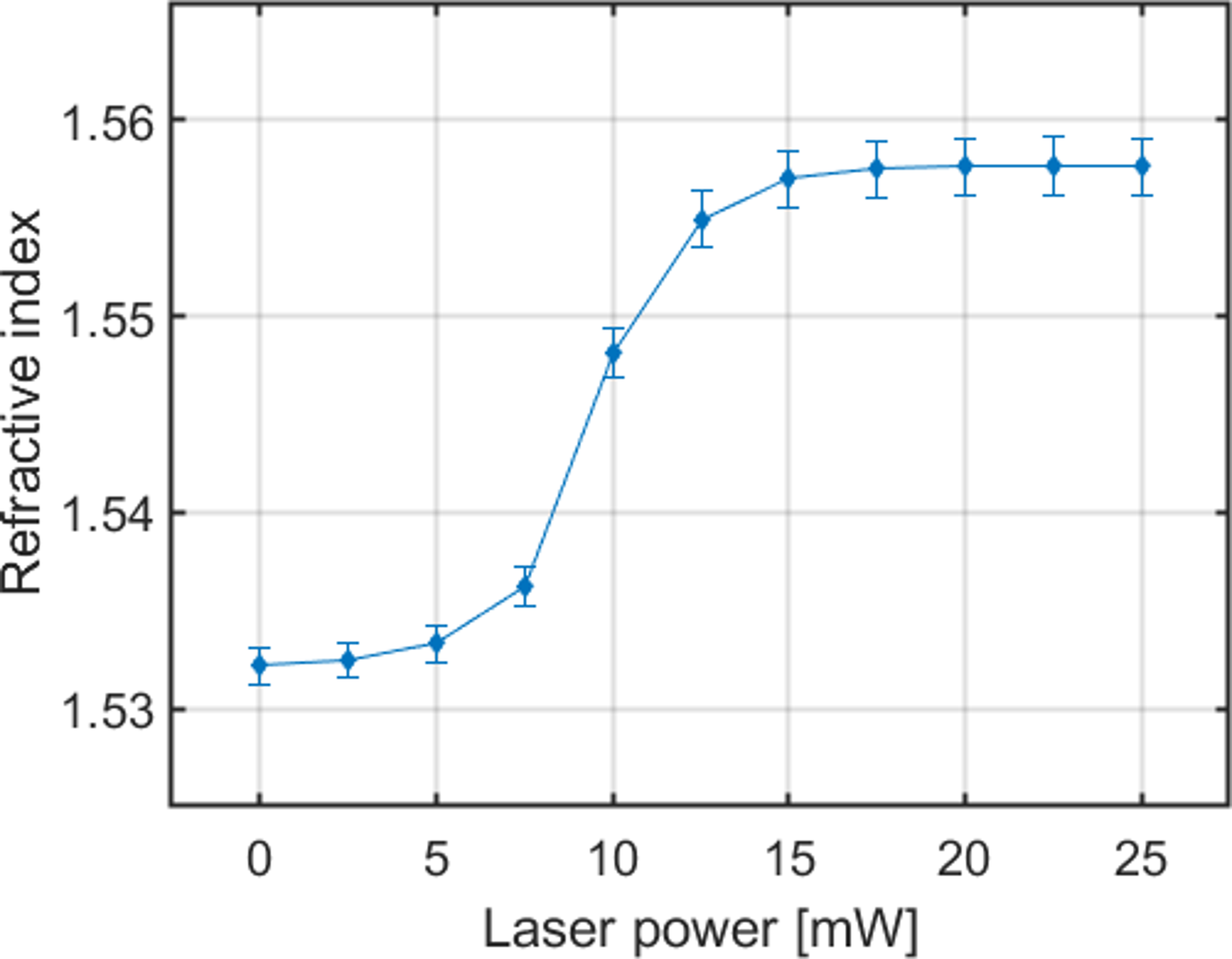} 
    \caption{Refractive index values of IP-Dip2 resin cured with various powers, calculated at \SI{633}{\nano \meter} wavelength and \SI{25}{\degreeCelsius}. Error bars represent expanded uncertainty (k=2).}
    \label{fig:ricalibration}
\end{figure}

RI calibration has been performed according to the methodology described in Sec.~\ref{sec:methodologyRIcalibration}. Laser power values in the resin chamber covered 0-60\% range with an interval of 5\%, and each power setting has been reprinted 4 times for a total of 52 individual steps. The RI measurements were performed 24 hours after fabrication to allow the resin to settle. The coverslip containing the calibration step structures was mounted in certified immersion liquid ($n_d$ = 1.50, Cargille Labs, USA) and covered with another coverslip prior to imaging in the DHM. A temperature probe was placed in contact with the sample and the temperature was noted at the first and last structure and averaged. After the first imaging set, the sample was unmounted, submerged in an isopropanol bath for 10 minutes, air-dried, mounted in the $n_d$ = 1.55 immersion liquid and measured again. Using the two immersion approach and the Eq.~\ref{eq:rimatching}, the calculated RI values as a function of the fabrication laser power are shown in Fig.~\ref{fig:ricalibration}. 

In order to validate whether the measurement and cleaning procedures are non-destructive, the sample was cleaned, mounted in $n_d$ = 1.50 medium and measured for a second time. Phase measurements obtained in $n_d$ = 1.55 and either the first $n_d$ = 1.50 (freshly mounted) or the repeated $n_d$ = 1.50 (twice remounted) provided two sets of values for the resin RI. The difference between those two sets is equal to 0.0002 on average, which is about 10x lower than the measurement uncertainty, with a standard deviation of 0.0002.

\subsection{Refractive index uncertainty} \label{sec:error} 
The uncertainty of the measured RI is calculated using the statistical deviations of the phase differences, as well as uncertainties of the RI medium and its temperature. In particular, the propagation of uncertainty is calculated using Eq.~\ref{eq:rimatching} and its partial derivatives with respect to each variable. 

In the case of phase uncertainty, first, for each step structure 50 phase images were collected and averaged. Temporal measurement precision was calculated as standard deviation of those 50 phases in each pixel. Then, the uncertainty of the phase of each ROI has been calculated as a geometrical mean of the precision map within corresponding ROIs. 

The uncertainty of the immersion media RI has been specified by their manufacturer. In particular, the Cargille Certified Immersion Liquids Series A were used with an expanded uncertainty of 0.0002, and their Cauchy and thermal coefficients were used to calculate the RI at the appropriate wavelength (\SI{633}{\nano \meter}) and temperature (in the range of 30--32\SI{}{\degreeCelsius} during measurements). 
The temperature of the sample chamber has been monitored using the DT-2 digital thermometer (Termoprodukt, Poland) with specified accuracy of \SI{0.1}{\degreeCelsius}. The temperature of a given measurement series is calculated as the average from readings noted at the first and the last structure. Since the temperature varied during the experiment even after initially achieving equilibrium, e.g. due to the heat generated by the mechanical stage used to reposition the sample, the temperature uncertainty used for calculations was increased to \SI{1}{\degreeCelsius}. 

For the data shown in Fig.~\ref{fig:ricalibration}, the expanded uncertainty (k = 2) ranges from 0.0009 to 0.0015. The increased uncertainty at higher DC is attributed mostly to the small RI difference between the polymer and the $n_d$ = 1.55 medium. The selection of immersion RIs could be further optimized to better balance the desired phase magnitude, noise and emergent uncertainty. In general, the phase uncertainty contributes only 10\% to the total uncertainty, while the immersion RI and temperature contribute about 30\% and 60\%, respectively. Structures for each power level were printed 4 times, so the repeatability of the measurement is calculated as the standard deviation of the RI values between those 4 steps. The RI repeatability turned out to be similar for all power levels and is equal to \SI{0.8e-4} on average.

For data unification and reporting, all RI values measured in this work have been recalculated to \SI{25}{\degreeCelsius}. The thermal coefficient for IP-Dip2 provided by the manufacturer is equal to \SI{-0.0004}{\per \degreeCelsius}. However, this coefficient might differ for liquid and solid states of the resin \cite{Feigl2025}, which would introduce an error that is not accounted for here.

\subsection{Refractive index aging} \label{sec:aging}
\begin{figure}[htbp]
    \includegraphics[width=.48\textwidth]{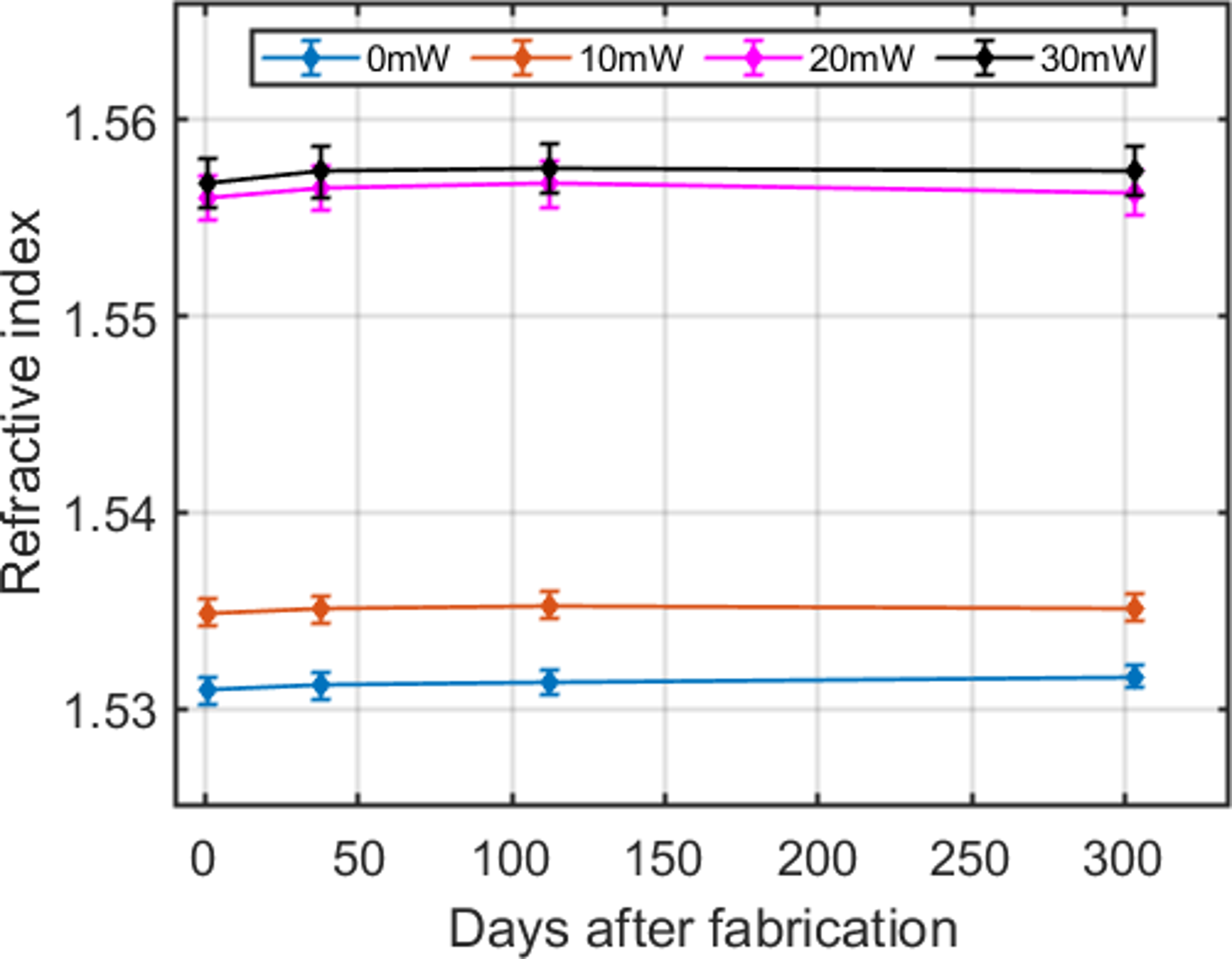} 
    \caption{Refractive index change of IP-Dip2 over time for four laser power values. The same sample has been measured 1, 38, 112 and 303 days after the fabrication. Error bars represent expanded uncertainty (k=2).}
    \label{fig:aging}
\end{figure}

The same methodology can be used to track the RI change in time due to aging. The RI values for four laser powers were measured 1, 38, 112 and 303 days after the fabrication and the results are shown in Fig.~\ref{fig:aging}. The highest peak-to-valley RI difference is 0.0007, recorded for the unprocessed resin on day 303, as well as for \SI{20}{\milli \watt} and \SI{30}{\milli \watt} dose on day 112. These differences are still below the measurement uncertainty and could be caused by random error. The average RI aging rate is \SI{5.66e-6} per day and there is no clear dependency on DC.

The sample has been stored in room temperature, wrapped in aluminium foil to protect from stray light and kept mounted in the immersion medium. These results highlight an important features of the resin for future deployment utilizing variable DC -- that it is quite stable under normal conditions and can be immersed in the RI-matching medium long-term.

Please note that the aging RI values are not directly comparable with Fig.~\ref{fig:ricalibration} as this sample has been printed using a different batch of resin and the hatching was slightly sparser, equal to \SI{0.2}{\micro \meter}.

\subsection{Voxel size calibration}
\begin{figure}[htbp]
    \includegraphics[width=.48\textwidth]{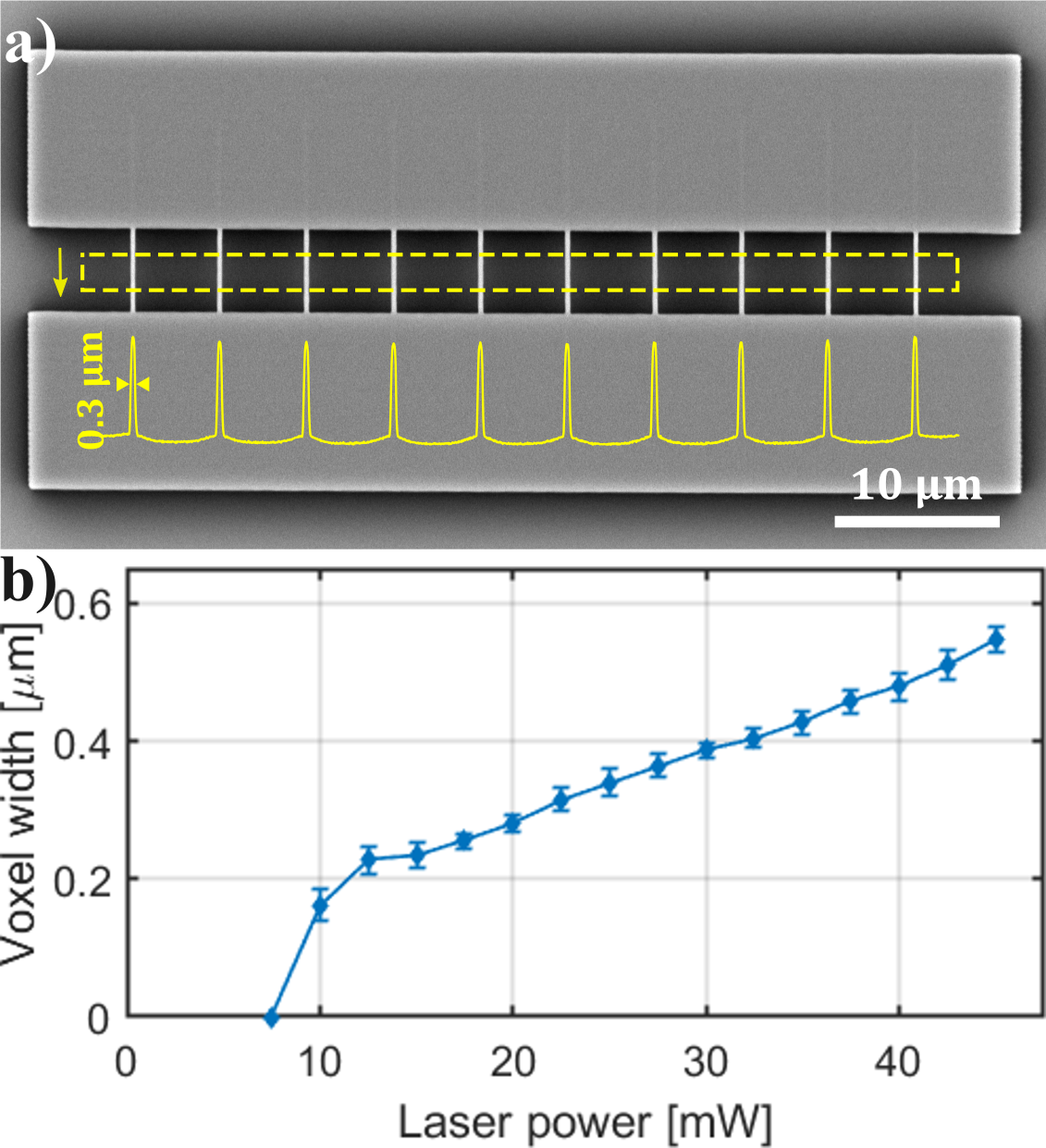} 
    \caption{Line width calibration of IP-Dip2 resin printed with various laser powers. a) Scanning electron microscopy image of suspended individual lines. Image area indicated by the dashed line is averaged in the direction indicated by the arrow to reduce noise, and then the width of the individual lines is evaluated at half maximum (inset). b) Voxel width calibration plot with the error bars indicating standard deviation from 10 measurements. }
    \label{fig:voxel_size}
\end{figure}

A similar calibration procedure has been performed with respect to the voxel size. A range of lines suspended in the air between two platforms were printed and their thickness measured using scanning electron microscopy. The results are presented in Fig.~\ref{fig:voxel_size}, where the results are averaged from 10 lines measured as full width at half maximum with the error bars indicating standard deviations. It is worth noting that the voxel size also depends on the distance from the center of the field of view due to additional optical aberrations of the off-axis beam. While printing close to the optical axis (in this case inside a radius of \SI{50}{\micro \meter}) this effect can be omitted. For larger fields it can also be calibrated and compensated -- see the "xycomp" option in the slicer implementation.

\subsection{3D refractive index structure design and fabrication}
\begin{figure*}[htbp]
    \centering
    \includegraphics[width=\textwidth]{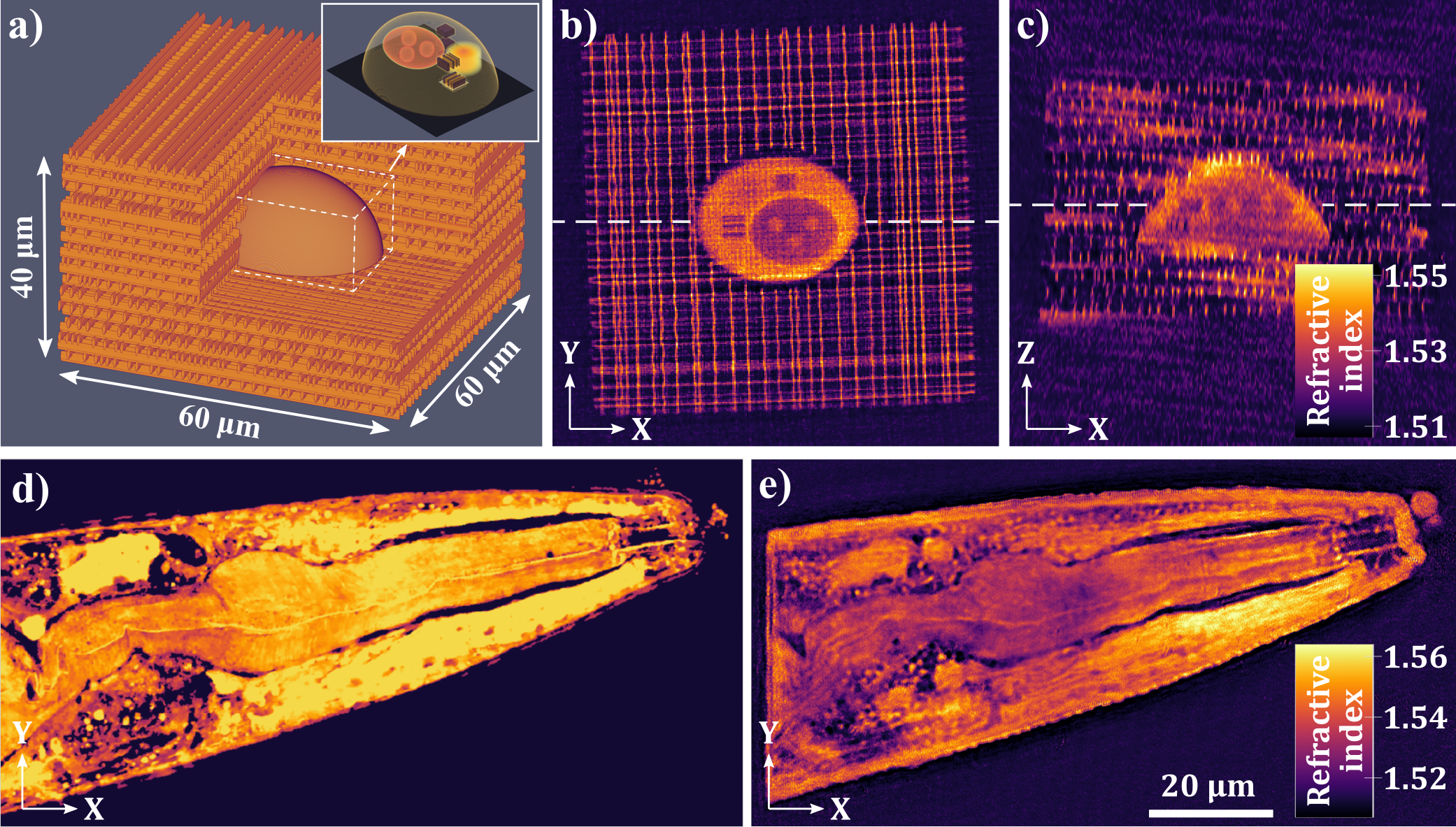} 
    \caption{3D refractive index engineering using the proposed slicer. a) Biological cell phantom design surrounded by the woodpile structure that introduces adjustable scattering.  b-c) Optical diffraction tomography reconstruction of the 3D RI of the cell target, shown as the XY and XZ cross-sections. d-e) Corresponding cross-sections of the design (d) and optical diffraction tomography reconstruction (e) of the printed C. elegans worm phantom, where all internal structures are visible purely due to the RI contrast modulated via printing laser power. Dataset for the C.~elegans worm is provided in the repository.}
    \label{fig:3dri}
\end{figure*}

The main utility of the slicer is streamlining the fabrication of structures with arbitrary DC distribution of the resin. Here it is demonstrated with tailor-made imaging targets having a 3D RI distribution for quantitative phase imaging applications, in particular for RI tomography. Fig.~\ref{fig:3dri}a) showcases the biological cell phantom -- an imaging target based on an eukariotic cell with visible nucleus, nucleoli and X/Y/Z resolution bars -- embedded within a woodpile structure intended to scatter light and hinder the imaging \cite{Krauze2022}. Such a design mimics the 3D RI distribution within real cells and allows to benchmark various imaging techniques while having full control over the scattering strength, adjusted via the size of the lattice. The experimental results shown in Fig.~\ref{fig:3dri}b-c) were obtained using the optical diffraction tomography (see \cite{Krauze2022} for details) and represent XY and XZ slices of the 3D RI reconstruction, confirming the successful fabrication of the overall geometry. The RI modulation within the cell is consistent with the designed distribution, as well as with the RI calibration shown in Sec.~\ref{sec:RIcalibration}. The overall design has been carefully crafted in MATLAB as a 3D RI volume and corresponding hatching map, directing the slicer to hatch e.g. individual lines along their length rather than width, which is also a crucial feature for resolution bars that are down to \SI{300}{\nano \meter} and must be printed in the direction perpendicular to their critical dimension.

Another example showcases the possibility of fabricating a 3D RI distribution from almost any source, be it measurement data (e.g. directly from 3D RI, rescaled confocal fluorescence or surface topology) or numerically optimized GRIN elements with a specialized function like beam shaping, coupling or diffraction. Here, the 3D RI reconstruction of the C. elegans worm published in \cite{Chowdhury2019} was preprocessed, printed and measured again. Fig.~\ref{fig:3dri}d) shows a single XY slice of the \SI{40}{\micro \meter} tall 3D RI design of the worm’s mouth, buccal cavity and the pharynx. Biological imaging results have been resampled and the RI values were rescaled to match the values addressable in the IP-Dip2 resin, decreasing the internal RI contrast by a factor of $\approx$2. It is worth noting, that these initially measured values can vary widely in terms of accuracy, however, the reprinted part should correspond well to the adjusted design as long as the proper RI calibration and structure integrity is ensured. Fig.~\ref{fig:3dri}e) shows the corresponding slice of the 3D RI reconstruction of the printed part, where all internal structures are visible purely due to the RI contrast modulated via the printing laser power. Nevertheless, some differences are still visible between these two cross-sections, and they are primarily due to the fact, that both TPP and limited-angle tomographic imaging have anisotropic resolution and suffer from elongation in the Z (axial) direction \cite{Zhou2023}. Therefore all features were printed taller than designed (which can be pre- or post-compensated for comparison), as well as the reconstructed 3D RI is further elongated in Z, resulting in out-of-plane crosstalk and general loss of the RI contrast. These results highlight the potential of the variable DC printing, as well as showcase extremely useful applications of TPP in metrology, where the phantom design, simulated reconstruction and experimental reconstruction can all be harnessed to quantify various types of errors. Such feedback and analysis is invaluable for benchmarking and evaluation of imaging systems, as well as some of their automated analysis pipelines for e.g. cell counting, segmentation or phenotyping. 

These two types of structures were printed with the same sampling (\qtyproduct[product-units = single]{0.15 x 0.15 x 0.3}{\micro\meter}) and they have similar volumes, however, exporting the files took less than a minute for the cell phantom and 20 minutes for the C. elegans sample on a medium-range workstation (AMD Ryzen 5 1600 CPU and 16GB of memory). The corresponding .gwl file sizes are 2 MB and 100 MB, which is of course correlated with the number of coordinates that need to be defined, as each laser power change along the line needs a new set of (x, y, z, laser power) values. So far, structure designs with high volume and entropy at file sizes above $\approx$500 MB occasionally cause errors of the printing software, however splitting the files into multiple Z-stitched chunks seems to solve this issue. Therefore, both the parsing time and these memory limitations are manageable even for high volume and demanding designs. 

\subsection{Grayscale lithography}
The relationship between the laser power and the voxel size can be used to perform grayscale lithography \cite{Aderneuer2021,Khonina2024}. In the structures printed with constant power only voxels that are within the designed geometry are printed. Using the grayscale approach, the residual height below the slicing value is addressed with either smaller voxels or taller voxels in the previous layer. Laser power values are programmatically adjusted to better approximate the height of the designed profile. 

\begin{figure}[htbp]
    \includegraphics[width=.48\textwidth]{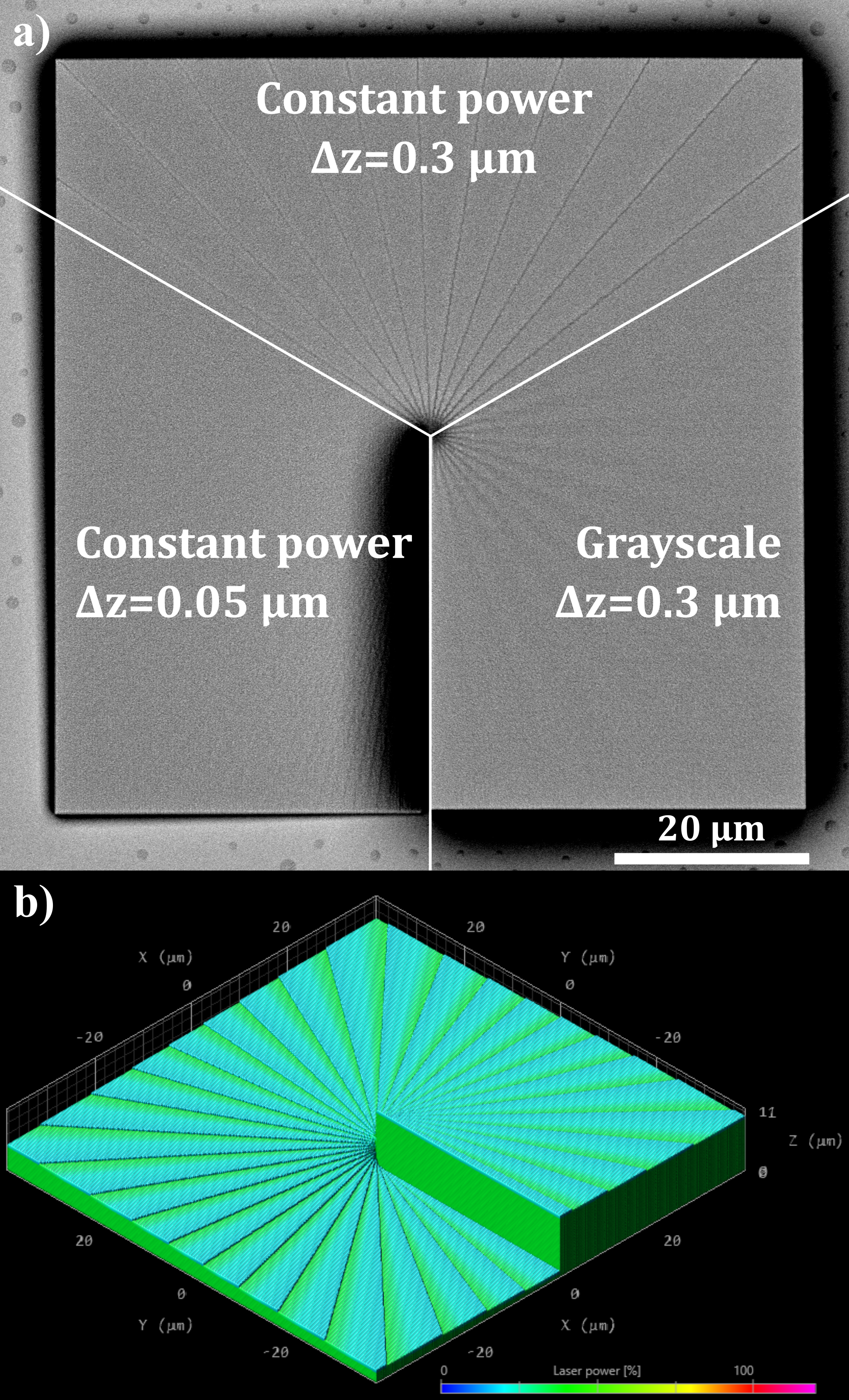} 
    \caption{Grayscale lithography demonstration using a \qtyproduct[product-units = single]{75 x 75 x 10}{\micro\meter} spiral phase plate. a) Scanning electron microscopy images of the three variants: printed with constant power (\SI{0.3}{\micro \meter} slicing), at 6x finer slicing (\SI{0.05}{\micro \meter}) and grayscale (\SI{0.3}{\micro \meter}). b) Laser power map of the grayscale design indicating automatic power adjustment to achieve accurate height profile.}
    \label{fig:grayscale}
\end{figure}

Experimental demonstration is shown in Fig.~\ref{fig:grayscale}, where the grayscale surface quality is comparable with 6x finer slicing (\SI{300}{\nano \meter} vs \SI{50}{\nano \meter}) without any additional print time. This result is achieved with an extremely simple algorithm defining grayscale structure, using 6-months-old voxel width calibration data and assuming that the voxel aspect ratio is 3x. Using other resins with larger processing window (like Nanoscribe's IP-S) and accurate voxel height calibration, the quality and speed uplift can be much greater, as evident by companies implementing this technique into their high-end systems. The code used to obtain these results is provided in the repository.

\subsection{Structure rotation} \label{sec:rotation}
Rotation of structures can be easily performed on .stl or voxelized design files. However, the slicer also enables to apply simple transformation matrices to the output coordinates of the print. This results in structures being rotated along with their hatching direction. Such global, post-hatching rotation could be useful to align printed structures to the features present on the substrate while also minimising any staircase effects in lateral direction when printing e.g. single lines or photonic crystal lattice. This feature was also utilized to fabricate the phantom embedded into the woodpile structure presented in Fig.~\ref{fig:3dri}a-c), which was printed in a single step from a single .gwl file, with each woodpile line and both X and Y resolution bars printed in their optimal orientation. Such structure rotation also enables to preserve critical dimensions while reducing drag of the structures during multimaterial printing. In this example, vacuum-based material exchange adapters making use of input and output nozzles, introduce resin flow to replace polymers in-situ, but perpendicular faces of the structures introduce stationary points \cite{Yemulwar2025}. These regions where flow is effectively zero can be minimized by structure rotation even if physical adjustment of the nozzle position is not possible. Ultimately, arbitrary transformation matrices can be implemented to control shear, twist or anisotropic shrinkage at the very end of the slicing process. 

\begin{figure}[htbp]
    \includegraphics[width=.48\textwidth]{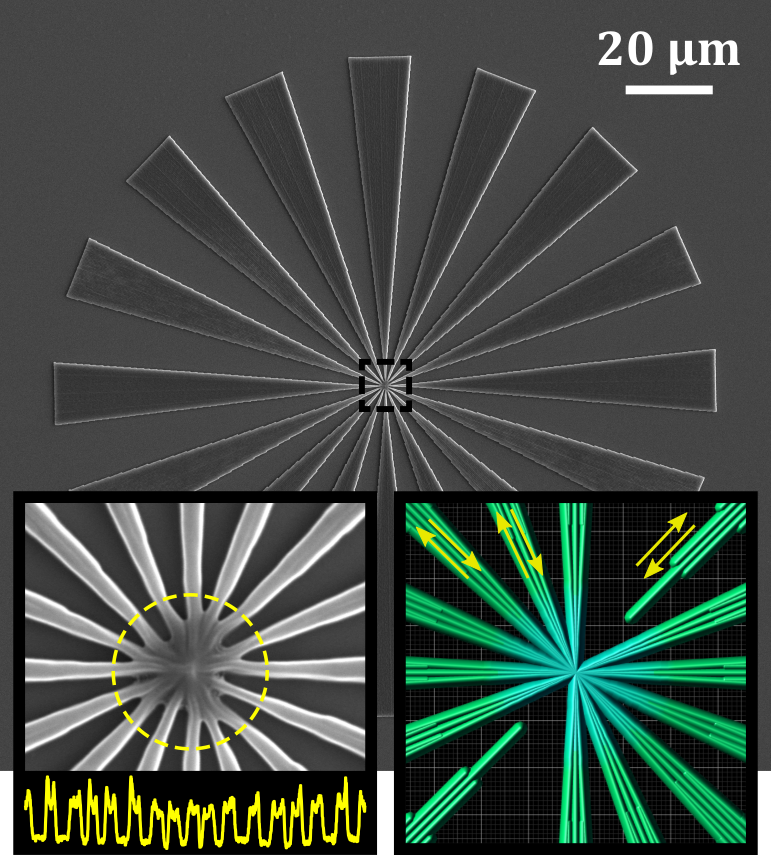} 
    \caption{Scanning electron microscopy image of a Siemens star fabricated with radial trajectories. Left inset: magnified view of the centerpoint. Line profile along the indicated circle (D=\SI{2.5}{\micro \meter}) corresponds to the \SI{250}{\nano \meter} half-pitch resolution. Right inset: rendered design during fabrication with indicated hatching directions and reduced laser power in the centerpoint.}
    \label{fig:siemens}
\end{figure}

One application example highlighting this feature is a Siemens star -- a popular resolution target with rotational symmetry \cite{Horstmeyer2016} that suffers from the staircase effect if printed with a fixed hatching. The optimized design consists of a single pair of triangles defined by two contour lines (edges) and infill hatching in the radial direction. The triangle pair is repeated (8 times in this case), each with appropriate rotation angle indicated in the slicer. The resulting structure is shown in Fig.~\ref{fig:siemens}, where $\approx$\SI{250}{\nano \meter} features are well-defined at \SI{1.5}{\micro \meter} total height, yielding an aspect ratio of 6. An additional dose adjustment in the centerpoint (shown in the bottom right inset) reduced the voxel size and dose accumulation, enabling at least \SI{250}{\nano \meter} half-pitch tangential resolution. An even smaller pitch should be achievable at lower structure heights and reduced surface tension during development.

\section{Discussion}
\subsection*{New opportunities for structure design and fabrication}
The new slicer proposed in this work relates to two major breakthroughs in TPP: design freedom and variable-DC fabrication. Structure definition no longer needs to adhere to restrictive mesh-based file formats, and instead, structures can be designed as voxelised power distribution maps that are native to the working principles of TPP. This enables a new class of structures that are designed and tailored for a particular task using e.g. finite element analysis, numerical and artificial intelligence-driven optimization or otherwise disordered layouts that rely not only on geometry of the part, but its core functions arise from the density heterogeneity. Several prominent variable-DC applications were already mentioned in the Introduction, but even fabrication on reflective substrates like silicon wafers benefit from a simple dose decrease for the first couple of layers to prevent burning. Variable DC incorporated into the microstructures opens the door for new opportunities and designs that were previously not feasible or prohibitorily complex in applications like actuators, self-assembling robots or metamaterials with exotic properties. The proposed slicer can replace existing solutions, such as programming the variable-DC structures directly in the .gwl or other low-level computer-aided manufacturing (CAM) languages, or splitting .stl files into several structures, each printed separately with different parameters. Presented solution enables workflows such as designing the structure's geometry in the user's preferred software, exporting it to an .stl file, voxelization and local adjustments to the printing power, and finally fabrication in a single job without any stitching or shadowing effects. 

In biomedical applications, designs derived from measurements are especially attractive. In-situ imaging such as confocal fluorescence can direct the polymerization beam to avoid exposing cells in cell-seeded structures. In such case, a fluorescence intensity map can serve as an inverse laser power map to reduce phototoxicity while maximising the dose in the scaffold, which lowers the number of unreacted, potentially cytotoxic functionalities \cite{VanHoorick2019}. Additionally, high DC reduce autofluorescence of the resin leading to enhanced contrast in confocal microscopy for cell biology applications \cite{Sharaf2023}, while adaptive and context-aware design \cite{Florczak2025} is utilized to automatically generate geometries that conform to various features on the substrate. All of these applications will be more accessible, integrated or further improved with the proposed slicer. 

Designs derived from measurements also support the development of digital twins for the TPP and imaging methods alike, where the numerical model of the system is continuously refined to better match predicted and observed output of the system. In TPP, for example \cite{Zheng2023} address the design-to-manufacturing gap by developing a photolithography simulator that compensates for errors introduced in the lithography process. Another example \cite{lvarezCastao2025} blends manufacturing and tomography by utilizing tomographic projections to induce volumetric polymerization. On the imaging side, 3D-printing is often utilized to create phantoms with specific shapes and contrast modalities \cite{Wang2017,Filippou2018,Horng2021,Ruiz2021}. TPP and phase imaging are particularly interesting due to their overlap and potential to address cell and tissue engineering metrology needs \cite{SimonJr2024}. Presented in Fig.~\ref{fig:3dri} print-outs and subsequent 3D RI reconstruction of the C. elegans, as well as previous works on microscopic bio-mimicking phantoms \cite{Micha2019,Krauze2022,Ziemczonok2025, Desissaire2025, Ziemczonok2022}, hint how well 3D RI fabrication and measurement complement each other already. Perfecting printing-measurement loop ensures symbiotic advancements for both fabrication and imaging techniques, and appropriate numerical tools and experimental expertise are essential for further developments in this space.

\subsection*{Refractive index calibration}
RI calibration plays a crucial role in the variable-DC printing as it connects the design of the microstructure with its desired behaviour. To date, many different approaches have been utilized to measure the RI, including goniometric or Pulfrich refractometers \cite{Gissibl2017,Ottermusch2019,Schmid2019,Feigl2025}, ellipsometry \cite{Li2019} or phase \cite{Zukauskas2015,Porte2021,Zvagelsky2023,Wdowiak2023,Hazem2025}, while other techniques related to Raman spectroscopy, shrinkage or stiffness measurements \cite{Jiang2014,LaFratta2017} have been used to quantify the DC. Unfortunately, none of the mentioned works could provide an appropriate results that would fully enable variable-DC printing, as they were obtained either for UV-cured resins, only for high DC (where the structure could withstand the development process) or at low or unspecified accuracy. The RI calibration methodology presented here requires custom TPP sample design (Fig.~\ref{fig:riramps}), but it enables non-destructive measurements of resin's RI for any DC, under practical fabrication conditions and with extremely high accuracy ($\approx0.1\%$) and precision ($\approx0.005\%$). 

\subsection*{Voxel, shrinkage and swelling pre- and post-compensation}
The slicer is designed primarily with the RI calibration in mind, providing strong metrological foundation for structure definition in order to boost confidence in the variable-DC fabrication approach. In principle, however, it is possible to use any calibration parameter to work out the necessary printing power distribution and such an example is also provided in the repository. Perhaps the most ubiquitous evaluation tool when working with TPP is scanning electron microscopy, so both calibration plots provided in Fig.~\ref{fig:ricalibration} (for RI) and Fig.~\ref{fig:voxel_size} (for voxel width) can be cross-referenced to make this work more accessible. Even though accurate calibration needs to be done for each material, microscope objective and even each setup due to misalignments and cleanliness, the general relationship between the voxel size and RI is useful to gauge the DC and to harness other, harder to measure quantities even if the RI data is not available.

The voxel size on its own is also an extremely powerful calibration parameter -- the 3D RI design of the microstructure can be augmented using the voxel size calibration to perform precompensation or obtain a more accurate prediction of the geometry or 3D RI from the print design. Both strategies can be implemented as simple erosion or dilation, respectively. As each designated dose creates a voxel of different size and RI value, the post-print compensation can be considered as a convolution of the designed power distribution with a power-adjusted voxel. Such a script is implemented, including proper overwrite rules where higher RI substitutes lower RI, and is also available in the repository. Furthermore, voxelised data format gives more pre- and post-compensation opportunities related to e.g. (anisotropic) shrinkage, swelling, anchor points or based on numerical simulations with respect to triggered responses to e.g. force or heat. Grayscale lithography can also be performed at any TPP system, which provides better surface quality and faster write times essentially for free. In the example shown in this work (Fig.~\ref{fig:grayscale}), only extremely basic calibration parameters were used -- 6-months-old voxel width data and the voxel height-to-width ratio of 3x. Direct voxel height calibration, as well as further hatching/slicing optimization can yield much better surface roughness and fabrication throughput. It further decouples the structure height profile from the stage Z-scanning, possibly enabling much higher quality of nanometer-scale features, critical in e.g. diffractive optical elements and optical processors. 

The open-source nature of the developed slicer enables further tuning of the printing process by the user, e.g. a dynamically adjustable hatching distance and/or writing direction (Sec.~\ref{sec:rotation}) can provide even greater control over the nanostructure of the part. Elliptical or even spiral scanning allows to further optimize the tools for the task and e.g. avoid stitching marks in critical locations. Full control over writing dynamics and the ability to quantify even tiny changes in material properties can shed new light on resin chemistry, proximity effects, oxygen diffusion or writing speed/time regimes \cite{Yang2019}, leading to more accurate numerical modelling of the TPP process and material science.

\subsection*{Challenges and limitations}
With all the advantages of variable-DC structures, they come with an obvious flaw -- the risk of uncontrolled photopolymerization due to stray light, which in the case of most popular resins include regular office light. As these structures are sensitive and fragile, they require special care and handling. We observed, that the dim table light and computer screen light do not alter the IP-Dip2 sample and in the span of a year it is otherwise stable (Sec.~\ref{sec:aging}), but depending on the material and use-case this can be a major limitation. Perhaps new strategies for stabilizing variable-DC regions or purging the remaining photoinitiators could alleviate this issue. 

Similarly to the stability of the resin, the stability and performance of the fabrication setup is crucial. The single-sample repeatability is high, but variations of temperature, humidity or dust build-up in the optical path of the system can render the calibration obsolete. Variable-DC is also demanding in terms of power switching speed and thus lower-end writing speeds are preferred. In this work the scanning speed was set to \SI{10000}{\micro \meter \per \second} with no visual errors at the single voxel scale (e.g. see tomographic reconstructions of the cell phantom in Fig.~\ref{fig:3dri}), but this should be further investigated and considered based on the structure design and requirements. 

Another limitation is related to the relatively small RI modulation range, in this case at the order of 0.03. For some applications and in other modalities this range can be too small, in which case an additional offset can be introduced e.g. with immersion medium. For higher gradients/ranges, other techniques appear better-suited, such as RI tuning by fill-factor of the polymerized resin within the porous scaffolds \cite{Ocier2020} or nanoparticles doping \cite{Ketchum2022,Prediger2024}.

Lastly, in terms of accessibility, the slicer is currently implemented only for one file format, with no guaranteed support for it from the TPP manufacturers. Exceptionally large files encounter some errors while loading them for the print, but fortunately they can be processed in smaller batches, which makes this approach fully-functional, even if less convenient. Hopefully, thanks to the open-source nature of the code and opportunities it creates, there will be an audience and incentive to provide support for variable-DC fabrication in the future and that it becomes a generic implementation into all commercial TPP platforms. 

\section{Conclusions}
This work provides an open-source slicer for variable-DC TPP fabrication, calibration methodology for determination of RI for any DC with extremely high accuracy and precision, and showcases various use-cases ranging from smoother surface quality and optimized imaging targets to bio-mimicking microorganisms phantoms. These tools and methods strive to popularize variable DC printing within the TPP community and to get more out of the existing TPP systems and workflows. Extensive RI measurement results demonstrate excellent repeatability, accuracy and stability of variable-DC structures and the strong metrological foundation builds confidence in the proposed approach. This work as a whole provides a complete toolbox for 3D refractive index engineering and sets the stage for new inventions enabled by point-wise dose control.

\section*{Data availability statement}
Supplementary data repository containing slicer source code, usage instructions, datasets and output .gwl files can be accessed at https://doi.org/10.5281/zenodo.18403965.

\section*{Acknowledgements}
The authors thank Prof. Shwetadwip Chowdhury from The University of Texas at Austin for providing the C.~elegans dataset.
\section*{Funding}
This work was supported by: 23IND08 DI-Vision grant by European Partnership on Metrology and EU Horizon Europe Research and Innovation Programme; the Warsaw University of Technology under the program Excellence Initiative: Research University (IDUB); Fonds Wetenschappelijk Onderzoek (12E2923N); Interreg (NWE758 “OIP4NWE”, “Fotonica Pilootlijnen”); Methusalem Foundation; IOF and OZR of Vrije Universiteit Brussel.

\section*{CRediT statement}
\textbf{MZ:}{Conceptualization, Methodology, Software, Investigation, Validation, Visualization, Writing - Original Draft.}
\textbf{KV:}{Investigation, Validation, Writing - Review \& Editing.}

\bibliography{bibliography}

\begin{thebibliography}{10}
\expandafter\ifx\csname url\endcsname\relax
  \def\url#1{\burl{#1}}\fi
\expandafter\ifx\csname urlprefix\endcsname\relax\def\urlprefix{URL }\fi
\providecommand{\bibinfo}[2]{#2}
\providecommand{\eprint}[2][]{\url{#2}}
\providecommand{\doi}[1]{\url{https://doi.org/#1}}
\bibcommenthead

\bibitem{Harinarayana2021}
\bibinfo{author}{Harinarayana, V.} \& \bibinfo{author}{Shin, Y.}
\newblock \bibinfo{title}{{Two-photon lithography for three-dimensional fabrication in micro/nanoscale regime: A comprehensive review}}.
\newblock \emph{\bibinfo{journal}{Optics \& Laser Technology}} \textbf{\bibinfo{volume}{142}}, \bibinfo{pages}{107180} (\bibinfo{year}{2021}).
\newblock \urlprefix\url{https://doi.org/10.1016/j.optlastec.2021.107180 https://linkinghub.elsevier.com/retrieve/pii/S0030399221002681}.

\bibitem{OHalloran2022}
\bibinfo{author}{O’Halloran, S.}, \bibinfo{author}{Pandit, A.}, \bibinfo{author}{Heise, A.} \& \bibinfo{author}{Kellett, A.}
\newblock \bibinfo{title}{Two‐photon polymerization: Fundamentals, materials, and chemical modification strategies}.
\newblock \emph{\bibinfo{journal}{Advanced Science}} \textbf{\bibinfo{volume}{10}} (\bibinfo{year}{2022}).
\newblock \urlprefix\url{http://dx.doi.org/10.1002/advs.202204072}.

\bibitem{Wang2024}
\bibinfo{author}{Wang, H.} \emph{et~al.}
\newblock \bibinfo{title}{{Two-photon polymerization lithography for imaging optics}}.
\newblock \emph{\bibinfo{journal}{International Journal of Extreme Manufacturing}} \textbf{\bibinfo{volume}{6}}, \bibinfo{pages}{042002} (\bibinfo{year}{2024}).
\newblock \urlprefix\url{https://iopscience.iop.org/article/10.1088/2631-7990/ad35fe}.

\bibitem{Hahn2020}
\bibinfo{author}{Hahn, V.} \emph{et~al.}
\newblock \bibinfo{title}{{Rapid Assembly of Small Materials Building Blocks (Voxels) into Large Functional 3D Metamaterials}}.
\newblock \emph{\bibinfo{journal}{Advanced Functional Materials}} \textbf{\bibinfo{volume}{30}} (\bibinfo{year}{2020}).
\newblock \urlprefix\url{https://advanced.onlinelibrary.wiley.com/doi/10.1002/adfm.201907795}.

\bibitem{LaFratta2017}
\bibinfo{author}{LaFratta, C.} \& \bibinfo{author}{Baldacchini, T.}
\newblock \bibinfo{title}{Two-photon polymerization metrology: Characterization methods of mechanisms and microstructures}.
\newblock \emph{\bibinfo{journal}{Micromachines}} \textbf{\bibinfo{volume}{8}}, \bibinfo{pages}{101} (\bibinfo{year}{2017}).
\newblock \urlprefix\url{http://dx.doi.org/10.3390/mi8040101}.

\bibitem{Vora2020}
\bibinfo{author}{Vora, H.~D.} \& \bibinfo{author}{Sanyal, S.}
\newblock \bibinfo{title}{{A comprehensive review: metrology in additive manufacturing and 3D printing technology}}.
\newblock \emph{\bibinfo{journal}{Progress in Additive Manufacturing}} \textbf{\bibinfo{volume}{5}}, \bibinfo{pages}{319--353} (\bibinfo{year}{2020}).
\newblock \urlprefix\url{https://doi.org/10.1007/s40964-020-00142-6 http://link.springer.com/10.1007/s40964-020-00142-6 https://link.springer.com/10.1007/s40964-020-00142-6}.

\bibitem{Carlotti2019}
\bibinfo{author}{Carlotti, M.} \& \bibinfo{author}{Mattoli, V.}
\newblock \bibinfo{title}{{Functional Materials for Two‐Photon Polymerization in Microfabrication}}.
\newblock \emph{\bibinfo{journal}{Small}} \textbf{\bibinfo{volume}{15}}, \bibinfo{pages}{1--22} (\bibinfo{year}{2019}).
\newblock \urlprefix\url{https://onlinelibrary.wiley.com/doi/10.1002/smll.201902687}.

\bibitem{Liao2020}
\bibinfo{author}{Liao, C.}, \bibinfo{author}{Wuethrich, A.} \& \bibinfo{author}{Trau, M.}
\newblock \bibinfo{title}{{A material odyssey for 3D nano/microstructures: two photon polymerization based nanolithography in bioapplications}}.
\newblock \emph{\bibinfo{journal}{Applied Materials Today}} \textbf{\bibinfo{volume}{19}}, \bibinfo{pages}{100635} (\bibinfo{year}{2020}).
\newblock \urlprefix\url{https://doi.org/10.1016/j.apmt.2020.100635 https://linkinghub.elsevier.com/retrieve/pii/S2352940720300834}.

\bibitem{Yang2021}
\bibinfo{author}{Yang, L.}, \bibinfo{author}{Mayer, F.}, \bibinfo{author}{Bunz, U. H.~F.}, \bibinfo{author}{Blasco, E.} \& \bibinfo{author}{Wegener, M.}
\newblock \bibinfo{title}{{Multi-material multi-photon 3D laser micro- and nanoprinting}}.
\newblock \emph{\bibinfo{journal}{Light: Advanced Manufacturing}} \textbf{\bibinfo{volume}{2}}, \bibinfo{pages}{1} (\bibinfo{year}{2021}).
\newblock \urlprefix\url{http://light-am.com/article/doi/10.37188/lam.2021.017 https://www.light-am.com/article/doi/10.37188/lam.2021.017}.

\bibitem{Jaiswal2023}
\bibinfo{author}{Jaiswal, A.} \emph{et~al.}
\newblock \bibinfo{title}{Two decades of two-photon lithography: Materials science perspective for additive manufacturing of 2d/3d nano-microstructures}.
\newblock \emph{\bibinfo{journal}{iScience}} \textbf{\bibinfo{volume}{26}}, \bibinfo{pages}{106374} (\bibinfo{year}{2023}).
\newblock \urlprefix\url{http://dx.doi.org/10.1016/j.isci.2023.106374}.

\bibitem{Song2024}
\bibinfo{author}{Song, D.} \emph{et~al.}
\newblock \bibinfo{title}{Generation of tailored multi‐material microstructures through one‐step direct laser writing}.
\newblock \emph{\bibinfo{journal}{Small}} \textbf{\bibinfo{volume}{20}} (\bibinfo{year}{2024}).
\newblock \urlprefix\url{http://dx.doi.org/10.1002/smll.202405586}.

\bibitem{ArriagaDvila2025}
\bibinfo{author}{Arriaga‐Dávila, J.} \emph{et~al.}
\newblock \bibinfo{title}{From single to multi‐material 3d printing of glass‐ceramics for micro‐optics}.
\newblock \emph{\bibinfo{journal}{Small Methods}} \textbf{\bibinfo{volume}{9}} (\bibinfo{year}{2025}).
\newblock \urlprefix\url{http://dx.doi.org/10.1002/smtd.202401809}.

\bibitem{Guney2016}
\bibinfo{author}{Guney, M.~G.} \& \bibinfo{author}{Fedder, G.~K.}
\newblock \bibinfo{title}{{Estimation of line dimensions in 3D direct laser writing lithography}}.
\newblock \emph{\bibinfo{journal}{Journal of Micromechanics and Microengineering}} \textbf{\bibinfo{volume}{26}}, \bibinfo{pages}{105011} (\bibinfo{year}{2016}).
\newblock \urlprefix\url{http://dx.doi.org/10.1088/0960-1317/26/10/105011}.

\bibitem{Bougdid2020}
\bibinfo{author}{Bougdid, Y.} \& \bibinfo{author}{Sekkat, Z.}
\newblock \bibinfo{title}{Voxels optimization in 3d laser nanoprinting}.
\newblock \emph{\bibinfo{journal}{Scientific Reports}} \textbf{\bibinfo{volume}{10}} (\bibinfo{year}{2020}).
\newblock \urlprefix\url{http://dx.doi.org/10.1038/s41598-020-67184-2}.

\bibitem{Drobecq2025}
\bibinfo{author}{Drobecq, I.}, \bibinfo{author}{Bigot, C.}, \bibinfo{author}{Soppera, O.}, \bibinfo{author}{Malaquin, L.} \& \bibinfo{author}{Venzac, B.}
\newblock \bibinfo{title}{Optimizing dimensional accuracy in two-photon polymerization: Influence of energy dose and proximity effects on sub-micrometric fiber structures}.
\newblock \emph{\bibinfo{journal}{Additive Manufacturing}} \textbf{\bibinfo{volume}{103}}, \bibinfo{pages}{104735} (\bibinfo{year}{2025}).
\newblock \urlprefix\url{http://dx.doi.org/10.1016/j.addma.2025.104735}.

\bibitem{Eren2025}
\bibinfo{author}{Eren, T.~N.} \emph{et~al.}
\newblock \bibinfo{title}{{Soft and Stiff 3D Microstructures by Step‐Growth Photopolymerization Using a Single Photoresin and Multi‐Photon Laser Printing}}.
\newblock \emph{\bibinfo{journal}{Advanced Functional Materials}} \textbf{\bibinfo{volume}{02876}}, \bibinfo{pages}{e02876} (\bibinfo{year}{2025}).
\newblock \urlprefix\url{/doi/pdf/10.1002/adfm.202502876%0Ahttps://onlinelibrary.wiley.com/doi/abs/10.1002/adfm.202502876%0Ahttps://advanced.onlinelibrary.wiley.com/doi/10.1002/adfm.202502876 https://advanced.onlinelibrary.wiley.com/doi/10.1002/adfm.202502876}.

\bibitem{Zukauskas2015}
\bibinfo{author}{{\v{Z}}ukauskas, A.} \emph{et~al.}
\newblock \bibinfo{title}{{Tuning the refractive index in 3D direct laser writing lithography: Towards GRIN microoptics}}.
\newblock \emph{\bibinfo{journal}{Laser and Photonics Reviews}} \textbf{\bibinfo{volume}{9}}, \bibinfo{pages}{706--712} (\bibinfo{year}{2015}).

\bibitem{Jiang2014}
\bibinfo{author}{Jiang, L.~J.} \emph{et~al.}
\newblock \bibinfo{title}{{Two-photon polymerization: investigation of chemical and mechanical properties of resins using Raman microspectroscopy}}.
\newblock \emph{\bibinfo{journal}{Optics Letters}} \textbf{\bibinfo{volume}{39}}, \bibinfo{pages}{3034} (\bibinfo{year}{2014}).

\bibitem{Sharaf2023}
\bibinfo{author}{Sharaf, A.}, \bibinfo{author}{Frimat, J.}, \bibinfo{author}{Kremers, G.} \& \bibinfo{author}{Accardo, A.}
\newblock \bibinfo{title}{{Suppression of auto-fluorescence from high-resolution 3D polymeric architectures fabricated via two-photon polymerization for cell biology applications}}.
\newblock \emph{\bibinfo{journal}{Micro and Nano Engineering}} \textbf{\bibinfo{volume}{19}}, \bibinfo{pages}{100188} (\bibinfo{year}{2023}).
\newblock \urlprefix\url{https://linkinghub.elsevier.com/retrieve/pii/S2590007223000187}.

\bibitem{Huang2020}
\bibinfo{author}{Huang, Z.}, \bibinfo{author}{{Chi-Pong Tsui}, G.}, \bibinfo{author}{Deng, Y.} \& \bibinfo{author}{Tang, C.-Y.}
\newblock \bibinfo{title}{{Two-photon polymerization nanolithography technology for fabrication of stimulus-responsive micro/nano-structures for biomedical applications}}.
\newblock \emph{\bibinfo{journal}{Nanotechnology Reviews}} \textbf{\bibinfo{volume}{9}}, \bibinfo{pages}{1118--1136} (\bibinfo{year}{2020}).
\newblock \urlprefix\url{https://www.degruyter.com/document/doi/10.1515/ntrev-2020-0073/html}.

\bibitem{Konara2025}
\bibinfo{author}{Konara, M.}, \bibinfo{author}{Pokharel, M.}, \bibinfo{author}{Sagar, M.~M.}, \bibinfo{author}{Kim, Y.} \& \bibinfo{author}{Park, K.}
\newblock \bibinfo{title}{Design, fabrication, and experimental validation of optical microbots}.
\newblock \emph{\bibinfo{journal}{Actuators}} \textbf{\bibinfo{volume}{14}}, \bibinfo{pages}{229} (\bibinfo{year}{2025}).
\newblock \urlprefix\url{http://dx.doi.org/10.3390/act14050229}.

\bibitem{Hippler2019}
\bibinfo{author}{Hippler, M.} \emph{et~al.}
\newblock \bibinfo{title}{{Controlling the shape of 3D microstructures by temperature and light}}.
\newblock \emph{\bibinfo{journal}{Nature Communications}} \textbf{\bibinfo{volume}{10}}, \bibinfo{pages}{232} (\bibinfo{year}{2019}).
\newblock \urlprefix\url{http://www.nature.com/articles/s41467-018-08175-w}.

\bibitem{Yarali2025}
\bibinfo{author}{Yarali, E.} \emph{et~al.}
\newblock \bibinfo{title}{Two-photon polymerization based 4d printing of poly(n-isopropylacrylamide) hydrogel microarchitectures for reversible shape morphing}.
\newblock \emph{\bibinfo{journal}{Scientific Reports}} \textbf{\bibinfo{volume}{15}} (\bibinfo{year}{2025}).
\newblock \urlprefix\url{http://dx.doi.org/10.1038/s41598-025-06269-2}.

\bibitem{Chen2025}
\bibinfo{author}{Chen, K.}, \bibinfo{author}{Thompson, A.~J.} \& \bibinfo{author}{Ahmad, B.}
\newblock \bibinfo{title}{Multi-dof optothermal microgripper for micromanipulation applications}.
\newblock \emph{\bibinfo{journal}{IEEE Robotics and Automation Letters}} \textbf{\bibinfo{volume}{10}}, \bibinfo{pages}{4061–4068} (\bibinfo{year}{2025}).
\newblock \urlprefix\url{http://dx.doi.org/10.1109/LRA.2025.3549240}.

\bibitem{Zhu2025}
\bibinfo{author}{Zhu, Y.}, \bibinfo{author}{Ma, Z.}, \bibinfo{author}{Sun, A.} \& \bibinfo{author}{Wei, Q.-H.}
\newblock \bibinfo{title}{Thermal responses of 3d-printed liquid crystal elastomer microcubes}.
\newblock \emph{\bibinfo{journal}{Liquid Crystals}} \bibinfo{pages}{1–8} (\bibinfo{year}{2025}).
\newblock \urlprefix\url{http://dx.doi.org/10.1080/02678292.2025.2499853}.

\bibitem{Salah2025}
\bibinfo{author}{Salah, M.} \emph{et~al.}
\newblock \bibinfo{title}{Towards bi-material 3d printed soft microrobots using two-photon polymerisation}.
\newblock \emph{\bibinfo{journal}{Journal of Micro and Bio Robotics}} \textbf{\bibinfo{volume}{21}} (\bibinfo{year}{2025}).
\newblock \urlprefix\url{http://dx.doi.org/10.1007/s12213-025-00185-4}.

\bibitem{Qu2017}
\bibinfo{author}{Qu, J.}, \bibinfo{author}{Kadic, M.}, \bibinfo{author}{Naber, A.} \& \bibinfo{author}{Wegener, M.}
\newblock \bibinfo{title}{{Micro-Structured Two-Component 3D Metamaterials with Negative Thermal-Expansion Coefficient from Positive Constituents}}.
\newblock \emph{\bibinfo{journal}{Scientific Reports}} \textbf{\bibinfo{volume}{7}}, \bibinfo{pages}{40643} (\bibinfo{year}{2017}).
\newblock \urlprefix\url{http://dx.doi.org/10.1038/srep40643 https://www.nature.com/articles/srep40643}.

\bibitem{VanHoorick2017}
\bibinfo{author}{Van~Hoorick, J.} \emph{et~al.}
\newblock \bibinfo{title}{Cross-linkable gelatins with superior mechanical properties through carboxylic acid modification: Increasing the two-photon polymerization potential}.
\newblock \emph{\bibinfo{journal}{Biomacromolecules}} \textbf{\bibinfo{volume}{18}}, \bibinfo{pages}{3260–3272} (\bibinfo{year}{2017}).
\newblock \urlprefix\url{http://dx.doi.org/10.1021/acs.biomac.7b00905}.

\bibitem{Ugarak2022}
\bibinfo{author}{Ugarak, F.} \emph{et~al.}
\newblock \bibinfo{title}{Brillouin light scattering characterisation of gray tone 3d printed isotropic materials}.
\newblock \emph{\bibinfo{journal}{Materials}} \textbf{\bibinfo{volume}{15}}, \bibinfo{pages}{4070} (\bibinfo{year}{2022}).
\newblock \urlprefix\url{http://dx.doi.org/10.3390/ma15124070}.

\bibitem{Aderneuer2021}
\bibinfo{author}{Aderneuer, T.}, \bibinfo{author}{Fernández, O.} \& \bibinfo{author}{Ferrini, R.}
\newblock \bibinfo{title}{Two-photon grayscale lithography for free-form micro-optical arrays}.
\newblock \emph{\bibinfo{journal}{Optics Express}} \textbf{\bibinfo{volume}{29}}, \bibinfo{pages}{39511} (\bibinfo{year}{2021}).
\newblock \urlprefix\url{http://dx.doi.org/10.1364/OE.440251}.

\bibitem{Khonina2024}
\bibinfo{author}{Khonina, S.~N.}, \bibinfo{author}{Kazanskiy, N.~L.} \& \bibinfo{author}{Butt, M.~A.}
\newblock \bibinfo{title}{{Grayscale Lithography and a Brief Introduction to Other Widely Used Lithographic Methods: A State-of-the-Art Review}}.
\newblock \emph{\bibinfo{journal}{Micromachines}} \textbf{\bibinfo{volume}{15}}, \bibinfo{pages}{1321} (\bibinfo{year}{2024}).
\newblock \urlprefix\url{https://www.mdpi.com/2072-666X/15/11/1321}.

\bibitem{Ye2025}
\bibinfo{author}{Ye, Y.} \emph{et~al.}
\newblock \bibinfo{title}{Microfabrication of hydrogels based on femtosecond laser three-dimensional ablation}.
\newblock \emph{\bibinfo{journal}{Chemical Engineering Journal}} \textbf{\bibinfo{volume}{520}}, \bibinfo{pages}{166034} (\bibinfo{year}{2025}).
\newblock \urlprefix\url{http://dx.doi.org/10.1016/j.cej.2025.166034}.

\bibitem{Ristok2020}
\bibinfo{author}{Ristok, S.}, \bibinfo{author}{Thiele, S.}, \bibinfo{author}{Toulouse, A.}, \bibinfo{author}{Herkommer, A.~M.} \& \bibinfo{author}{Giessen, H.}
\newblock \bibinfo{title}{Stitching-free 3d printing of millimeter-sized highly transparent spherical and aspherical optical components}.
\newblock \emph{\bibinfo{journal}{Optical Materials Express}} \textbf{\bibinfo{volume}{10}}, \bibinfo{pages}{2370} (\bibinfo{year}{2020}).
\newblock \urlprefix\url{http://dx.doi.org/10.1364/OME.401724}.

\bibitem{GonzalezHernandez2021}
\bibinfo{author}{Gonzalez-Hernandez, D.} \emph{et~al.}
\newblock \bibinfo{title}{Laser 3d printing of inorganic free-form micro-optics}.
\newblock \emph{\bibinfo{journal}{Photonics}} \textbf{\bibinfo{volume}{8}}, \bibinfo{pages}{577} (\bibinfo{year}{2021}).
\newblock \urlprefix\url{http://dx.doi.org/10.3390/photonics8120577}.

\bibitem{Liu2025}
\bibinfo{author}{Liu, J.}, \bibinfo{author}{Lycke, R.}, \bibinfo{author}{Luan, L.} \& \bibinfo{author}{Tkaczyk, T.~S.}
\newblock \bibinfo{title}{Fully 3d-printed endomicroscopic objective for two-photon, multi-wavelength excitation microscopy}.
\newblock \emph{\bibinfo{journal}{Biomedical Optics Express}} \textbf{\bibinfo{volume}{17}}, \bibinfo{pages}{82} (\bibinfo{year}{2025}).
\newblock \urlprefix\url{http://dx.doi.org/10.1364/BOE.579476}.

\bibitem{GonzalezHernandez2023}
\bibinfo{author}{Gonzalez‐Hernandez, D.} \emph{et~al.}
\newblock \bibinfo{title}{{Single‐Step 3D Printing of Micro‐Optics with Adjustable Refractive Index by Ultrafast Laser Nanolithography}}.
\newblock \emph{\bibinfo{journal}{Advanced Optical Materials}} \textbf{\bibinfo{volume}{2300258}}, \bibinfo{pages}{1--7} (\bibinfo{year}{2023}).
\newblock \urlprefix\url{https://onlinelibrary.wiley.com/doi/10.1002/adom.202300258}.

\bibitem{Porte2021}
\bibinfo{author}{Porte, X.} \emph{et~al.}
\newblock \bibinfo{title}{{Direct (3+1)D laser writing of graded-index optical elements}}.
\newblock \emph{\bibinfo{journal}{Optica}} \textbf{\bibinfo{volume}{8}}, \bibinfo{pages}{1281} (\bibinfo{year}{2021}).
\newblock \urlprefix\url{https://opg.optica.org/abstract.cfm?URI=optica-8-10-1281}.

\bibitem{Micha2019}
\bibinfo{author}{Ziemczonok, M.}, \bibinfo{author}{Ku{\'{s}}, A.}, \bibinfo{author}{Wasylczyk, P.} \& \bibinfo{author}{Kujawi{\'{n}}ska, M.}
\newblock \bibinfo{title}{{3D-printed biological cell phantom for testing 3D quantitative phase imaging systems}}.
\newblock \emph{\bibinfo{journal}{Scientific Reports}} \textbf{\bibinfo{volume}{9}}, \bibinfo{pages}{18872} (\bibinfo{year}{2019}).
\newblock \urlprefix\url{http://www.nature.com/articles/s41598-019-55330-4}.

\bibitem{Lamont2020}
\bibinfo{author}{Lamont, A.~C.} \emph{et~al.}
\newblock \bibinfo{title}{{Direct laser writing of a titanium dioxide-laden retinal cone phantom for adaptive optics-optical coherence tomography}}.
\newblock \emph{\bibinfo{journal}{Optical Materials Express}} \textbf{\bibinfo{volume}{10}}, \bibinfo{pages}{2757} (\bibinfo{year}{2020}).
\newblock \urlprefix\url{https://opg.optica.org/abstract.cfm?URI=ome-10-11-2757}.

\bibitem{Krauze2022}
\bibinfo{author}{Krauze, W.} \emph{et~al.}
\newblock \bibinfo{title}{{3D scattering microphantom sample to assess quantitative accuracy in tomographic phase microscopy techniques}}.
\newblock \emph{\bibinfo{journal}{Scientific Reports}} \textbf{\bibinfo{volume}{12}}, \bibinfo{pages}{19586} (\bibinfo{year}{2022}).
\newblock \urlprefix\url{http://arxiv.org/abs/2208.09311 https://www.nature.com/articles/s41598-022-24193-7}.

\bibitem{Kim2025}
\bibinfo{author}{Kim, J.} \emph{et~al.}
\newblock \bibinfo{title}{{Inverse-scattering in biological samples via beam-propagation}} (\bibinfo{year}{2025}).
\newblock \urlprefix\url{http://biorxiv.org/lookup/doi/10.1101/2025.08.17.670744}.

\bibitem{Ziemczonok2025}
\bibinfo{author}{Ziemczonok, M.} \emph{et~al.}
\newblock \bibinfo{title}{{Tailored 3D microphantoms: An essential tool for quantitative phase tomography analysis of organoids}}.
\newblock \emph{\bibinfo{journal}{Biocybernetics and Biomedical Engineering}} \textbf{\bibinfo{volume}{45}}, \bibinfo{pages}{247--257} (\bibinfo{year}{2025}).
\newblock \urlprefix\url{https://doi.org/10.1016/j.bbe.2025.03.003 https://linkinghub.elsevier.com/retrieve/pii/S0208521625000221}.

\bibitem{Kim2011}
\bibinfo{author}{Kim, M.~K.}
\newblock \emph{\bibinfo{title}{Digital Holographic Microscopy: Principles, Techniques, and Applications}}  (\bibinfo{publisher}{Springer New York}, \bibinfo{year}{2011}).
\newblock \urlprefix\url{http://dx.doi.org/10.1007/978-1-4419-7793-9}.

\bibitem{Besaga2019}
\bibinfo{author}{Besaga, V.~R.}, \bibinfo{author}{Saetchnikov, A.~V.}, \bibinfo{author}{Gerhardt, N.~C.}, \bibinfo{author}{Ostendorf, A.} \& \bibinfo{author}{Hofmann, M.~R.}
\newblock \bibinfo{title}{{Digital holographic microscopy for sub-µm scale high aspect ratio structures in transparent materials}}.
\newblock \emph{\bibinfo{journal}{Optics and Lasers in Engineering}} \textbf{\bibinfo{volume}{121}}, \bibinfo{pages}{441--447} (\bibinfo{year}{2019}).
\newblock \urlprefix\url{https://doi.org/10.1016/j.optlaseng.2019.05.007 https://linkinghub.elsevier.com/retrieve/pii/S0143816618316269}.

\bibitem{Emery2021}
\bibinfo{author}{Emery, Y.}, \bibinfo{author}{Colomb, T.} \& \bibinfo{author}{Cuche, E.}
\newblock \bibinfo{title}{{Metrology applications using off-axis digital holography microscopy}}.
\newblock \emph{\bibinfo{journal}{Journal of Physics: Photonics}} \textbf{\bibinfo{volume}{3}}, \bibinfo{pages}{034016} (\bibinfo{year}{2021}).
\newblock \urlprefix\url{https://iopscience.iop.org/article/10.1088/2515-7647/ac0957}.

\bibitem{Chaumet2024}
\bibinfo{author}{Chaumet, P.~C.}, \bibinfo{author}{Bon, P.}, \bibinfo{author}{Maire, G.}, \bibinfo{author}{Sentenac, A.} \& \bibinfo{author}{Baffou, G.}
\newblock \bibinfo{title}{{Quantitative phase microscopies: accuracy comparison}}.
\newblock \emph{\bibinfo{journal}{Light: Science \& Applications}} \textbf{\bibinfo{volume}{13}}, \bibinfo{pages}{288} (\bibinfo{year}{2024}).
\newblock \urlprefix\url{http://arxiv.org/abs/2403.11930 https://www.nature.com/articles/s41377-024-01619-7}.

\bibitem{Desissaire2025}
\bibinfo{author}{Desissaire, S.} \emph{et~al.}
\newblock \bibinfo{title}{{Bio-inspired 3D-printed phantom: Encoding cellular heterogeneity for characterization of quantitative phase imaging}}.
\newblock \emph{\bibinfo{journal}{Measurement}} \textbf{\bibinfo{volume}{247}}, \bibinfo{pages}{116765} (\bibinfo{year}{2025}).
\newblock \urlprefix\url{https://linkinghub.elsevier.com/retrieve/pii/S0263224125001241}.

\bibitem{stl_FEX}
\bibinfo{author}{Micó, P.}
\newblock \bibinfo{title}{stltools}.
\newblock \bibinfo{howpublished}{\url{https://www.mathworks.com/matlabcentral/fileexchange/51200-stltools}} (\bibinfo{year}{2017}).
\newblock \bibinfo{note}{[Online; accessed December 18, 2025]}.

\bibitem{Feigl2025}
\bibinfo{author}{Feigl, G.}, \bibinfo{author}{Zaugg, D.}, \bibinfo{author}{Hinum-Wagner, J.~W.}, \bibinfo{author}{H\"{o}rmann, S.~M.} \& \bibinfo{author}{Bergmann, A.}
\newblock \bibinfo{title}{Refractive index characterization and modeling of polymerization-dependent refractive index variations in two-photon polymerization resins}.
\newblock \emph{\bibinfo{journal}{Optics Continuum}} \textbf{\bibinfo{volume}{4}}, \bibinfo{pages}{996} (\bibinfo{year}{2025}).
\newblock \urlprefix\url{http://dx.doi.org/10.1364/OPTCON.557710}.

\bibitem{Chowdhury2019}
\bibinfo{author}{Chowdhury, S.} \emph{et~al.}
\newblock \bibinfo{title}{{High-resolution 3D refractive index microscopy of multiple-scattering samples from intensity images}}.
\newblock \emph{\bibinfo{journal}{Optica}} \textbf{\bibinfo{volume}{6}}, \bibinfo{pages}{1211} (\bibinfo{year}{2019}).
\newblock \urlprefix\url{https://www.osapublishing.org/abstract.cfm?URI=optica-6-9-1211}.

\bibitem{Zhou2023}
\bibinfo{author}{Zhou, N.} \emph{et~al.}
\newblock \bibinfo{title}{{Quasi-Isotropic High-Resolution Fourier Ptychographic Diffraction Tomography with Opposite Illuminations}}.
\newblock \emph{\bibinfo{journal}{ACS Photonics}} \textbf{\bibinfo{volume}{10}}, \bibinfo{pages}{2461--2466} (\bibinfo{year}{2023}).
\newblock \urlprefix\url{https://pubs.acs.org/doi/10.1021/acsphotonics.3c00227}.

\bibitem{Yemulwar2025}
\bibinfo{author}{Yemulwar, P.} \emph{et~al.}
\newblock \bibinfo{editor}{Helvajian, H.}, \bibinfo{editor}{Gu, B.} \& \bibinfo{editor}{Chen, H.} (eds) \emph{\bibinfo{title}{Flow dynamics of multi-material exchange in two-photon absorption 3d printing}}.
\newblock (eds \bibinfo{editor}{Helvajian, H.}, \bibinfo{editor}{Gu, B.} \& \bibinfo{editor}{Chen, H.}) \emph{\bibinfo{booktitle}{Laser 3D Manufacturing XII}}, \bibinfo{pages}{1} (\bibinfo{publisher}{SPIE}, \bibinfo{year}{2025}).
\newblock \urlprefix\url{http://dx.doi.org/10.1117/12.3040656}.

\bibitem{Horstmeyer2016}
\bibinfo{author}{Horstmeyer, R.}, \bibinfo{author}{Heintzmann, R.}, \bibinfo{author}{Popescu, G.}, \bibinfo{author}{Waller, L.} \& \bibinfo{author}{Yang, C.}
\newblock \bibinfo{title}{{Standardizing the resolution claims for coherent microscopy}}.
\newblock \emph{\bibinfo{journal}{Nature Photonics}} \textbf{\bibinfo{volume}{10}}, \bibinfo{pages}{68--71} (\bibinfo{year}{2016}).
\newblock \urlprefix\url{http://dx.doi.org/10.1038/nphoton.2015.279 http://www.nature.com/articles/nphoton.2015.279}.

\bibitem{VanHoorick2019}
\bibinfo{author}{Van~Hoorick, J.} \emph{et~al.}
\newblock \bibinfo{title}{(photo-)crosslinkable gelatin derivatives for biofabrication applications}.
\newblock \emph{\bibinfo{journal}{Acta Biomaterialia}} \textbf{\bibinfo{volume}{97}}, \bibinfo{pages}{46–73} (\bibinfo{year}{2019}).
\newblock \urlprefix\url{http://dx.doi.org/10.1016/j.actbio.2019.07.035}.

\bibitem{Florczak2025}
\bibinfo{author}{Florczak, S.} \emph{et~al.}
\newblock \bibinfo{title}{Adaptive and context-aware volumetric printing}.
\newblock \emph{\bibinfo{journal}{Nature}} \textbf{\bibinfo{volume}{645}}, \bibinfo{pages}{108–114} (\bibinfo{year}{2025}).
\newblock \urlprefix\url{http://dx.doi.org/10.1038/s41586-025-09436-7}.

\bibitem{Zheng2023}
\bibinfo{author}{Zheng, C.}, \bibinfo{author}{Zhao, G.} \& \bibinfo{author}{So, P.}
\newblock \bibinfo{title}{Close the design-to-manufacturing gap in computational optics with a 'real2sim' learned two-photon neural lithography simulator} (\bibinfo{year}{2023}).
\newblock \urlprefix\url{https://doi.org/10.1145/3610548.3618251}.

\bibitem{lvarezCastao2025}
\bibinfo{author}{Álvarez Castaño, M.~I.} \emph{et~al.}
\newblock \bibinfo{title}{Holographic tomographic volumetric additive manufacturing}.
\newblock \emph{\bibinfo{journal}{Nature Communications}} \textbf{\bibinfo{volume}{16}} (\bibinfo{year}{2025}).
\newblock \urlprefix\url{http://dx.doi.org/10.1038/s41467-025-56852-4}.

\bibitem{Wang2017}
\bibinfo{author}{Wang, K.}, \bibinfo{author}{Ho, C.-C.}, \bibinfo{author}{Zhang, C.} \& \bibinfo{author}{Wang, B.}
\newblock \bibinfo{title}{{A Review on the 3D Printing of Functional Structures for Medical Phantoms and Regenerated Tissue and Organ Applications}}.
\newblock \emph{\bibinfo{journal}{Engineering}} \textbf{\bibinfo{volume}{3}}, \bibinfo{pages}{653--662} (\bibinfo{year}{2017}).
\newblock \urlprefix\url{http://dx.doi.org/10.1016/J.ENG.2017.05.013 https://linkinghub.elsevier.com/retrieve/pii/S2095809917307178}.

\bibitem{Filippou2018}
\bibinfo{author}{Filippou, V.} \& \bibinfo{author}{Tsoumpas, C.}
\newblock \bibinfo{title}{{Recent advances on the development of phantoms using 3D printing for imaging with CT, MRI, PET, SPECT, and ultrasound}}.
\newblock \emph{\bibinfo{journal}{Medical Physics}} \textbf{\bibinfo{volume}{45}}, \bibinfo{pages}{e740--e760} (\bibinfo{year}{2018}).
\newblock \urlprefix\url{http://doi.wiley.com/10.1002/mp.13058}.

\bibitem{Horng2021}
\bibinfo{author}{Horng, H.} \emph{et~al.}
\newblock \bibinfo{title}{{3D printed vascular phantoms for high-resolution biophotonic image quality assessment via direct laser writing}}.
\newblock \emph{\bibinfo{journal}{Optics Letters}} \textbf{\bibinfo{volume}{46}}, \bibinfo{pages}{1987} (\bibinfo{year}{2021}).

\bibitem{Ruiz2021}
\bibinfo{author}{Ruiz, A.~J.} \emph{et~al.}
\newblock \bibinfo{title}{{3D printing fluorescent material with tunable optical properties}}.
\newblock \emph{\bibinfo{journal}{Scientific Reports}} \textbf{\bibinfo{volume}{11}}, \bibinfo{pages}{17135} (\bibinfo{year}{2021}).
\newblock \urlprefix\url{https://doi.org/10.1038/s41598-021-96496-0 https://www.nature.com/articles/s41598-021-96496-0}.

\bibitem{SimonJr2024}
\bibinfo{author}{Babakhanova, G.}, \bibinfo{author}{Simon, C.~G.} \& \bibinfo{author}{Romantseva, E.}
\newblock \emph{\bibinfo{title}{Measurement needs for biofabrication of tissue engineered medical products workshop report}}  (\bibinfo{publisher}{National Institute of Standards and Technology}, \bibinfo{year}{2024}).
\newblock \urlprefix\url{http://dx.doi.org/10.6028/NIST.SP.1500-23}.

\bibitem{Ziemczonok2022}
\bibinfo{author}{Ziemczonok, M.}, \bibinfo{author}{Ku{\'{s}}, A.} \& \bibinfo{author}{Kujawi{\'{n}}ska, M.}
\newblock \bibinfo{title}{{Optical diffraction tomography meets metrology — Measurement accuracy on cellular and subcellular level}}.
\newblock \emph{\bibinfo{journal}{Measurement}} \textbf{\bibinfo{volume}{195}}, \bibinfo{pages}{111106} (\bibinfo{year}{2022}).
\newblock \urlprefix\url{https://doi.org/10.1016/j.measurement.2022.111106 https://linkinghub.elsevier.com/retrieve/pii/S0263224122003700}.

\bibitem{Gissibl2017}
\bibinfo{author}{Gissibl, T.}, \bibinfo{author}{Wagner, S.}, \bibinfo{author}{Sykora, J.}, \bibinfo{author}{Schmid, M.} \& \bibinfo{author}{Giessen, H.}
\newblock \bibinfo{title}{{Refractive index measurements of photo-resists for three-dimensional direct laser writing}}.
\newblock \emph{\bibinfo{journal}{Optical Materials Express}} \textbf{\bibinfo{volume}{7}}, \bibinfo{pages}{2293} (\bibinfo{year}{2017}).
\newblock \urlprefix\url{https://www.osapublishing.org/abstract.cfm?URI=ome-7-7-2293 https://link.springer.com/10.1007/BF00760845}.

\bibitem{Ottermusch2019}
\bibinfo{author}{Dottermusch, S.}, \bibinfo{author}{Busko, D.}, \bibinfo{author}{Langenhorst, M.}, \bibinfo{author}{Paetzold, U.~W.} \& \bibinfo{author}{Richards, B.~S.}
\newblock \bibinfo{title}{{Exposure-dependent refractive index of Nanoscribe IP-Dip photoresist layers}}.
\newblock \emph{\bibinfo{journal}{Optics Letters}} \textbf{\bibinfo{volume}{44}}, \bibinfo{pages}{29} (\bibinfo{year}{2019}).
\newblock \urlprefix\url{https://opg.optica.org/abstract.cfm?URI=ol-44-1-29}.

\bibitem{Schmid2019}
\bibinfo{author}{Schmid, M.}, \bibinfo{author}{Ludescher, D.} \& \bibinfo{author}{Giessen, H.}
\newblock \bibinfo{title}{{Optical properties of photoresists for femtosecond 3D printing: refractive index, extinction, luminescence-dose dependence, aging, heat treatment and comparison between 1-photon and 2-photon exposure}}.
\newblock \emph{\bibinfo{journal}{Optical Materials Express}} \textbf{\bibinfo{volume}{9}}, \bibinfo{pages}{4564} (\bibinfo{year}{2019}).
\newblock \urlprefix\url{https://www.osapublishing.org/abstract.cfm?URI=ome-9-12-4564 https://opg.optica.org/abstract.cfm?URI=ome-9-12-4564}.

\bibitem{Li2019}
\bibinfo{author}{Li, Y.} \emph{et~al.}
\newblock \bibinfo{title}{Uv to nir optical properties of ip-dip, ip-l, and ip-s after two-photon polymerization determined by spectroscopic ellipsometry}.
\newblock \emph{\bibinfo{journal}{Optical Materials Express}} \textbf{\bibinfo{volume}{9}}, \bibinfo{pages}{4318} (\bibinfo{year}{2019}).
\newblock \urlprefix\url{http://dx.doi.org/10.1364/OME.9.004318}.

\bibitem{Zvagelsky2023}
\bibinfo{author}{Zvagelsky, R.}, \bibinfo{author}{Kiefer, P.}, \bibinfo{author}{Weinacker, J.} \& \bibinfo{author}{Wegener, M.}
\newblock \bibinfo{title}{{In-situ Quantitative Phase Imaging during Multi-photon Laser Printing}}.
\newblock \emph{\bibinfo{journal}{ACS Photonics}}  (\bibinfo{year}{2023}).
\newblock \urlprefix\url{https://pubs.acs.org/doi/10.1021/acsphotonics.3c00625}.

\bibitem{Wdowiak2023}
\bibinfo{author}{Wdowiak, E.}, \bibinfo{author}{Ziemczonok, M.}, \bibinfo{author}{Martinez-Carranza, J.} \& \bibinfo{author}{Ku{\'{s}}, A.}
\newblock \bibinfo{title}{{Phase-assisted multi-material two-photon polymerization for extended refractive index range}}.
\newblock \emph{\bibinfo{journal}{Additive Manufacturing}} \textbf{\bibinfo{volume}{73}}, \bibinfo{pages}{103666} (\bibinfo{year}{2023}).
\newblock \urlprefix\url{https://doi.org/10.1016/j.addma.2023.103666 https://linkinghub.elsevier.com/retrieve/pii/S2214860423002798}.

\bibitem{Hazem2025}
\bibinfo{author}{Hazem, R.}, \bibinfo{author}{Carpentier, M.}, \bibinfo{author}{Dussauze, M.}, \bibinfo{author}{Petit, Y.} \& \bibinfo{author}{Canioni, L.}
\newblock \bibinfo{title}{{(3+1)D Printing of Core–Clad Waveguide by Two‐Photon Polymerization}}.
\newblock \emph{\bibinfo{journal}{Advanced Materials Technologies}} \textbf{\bibinfo{volume}{10}}, \bibinfo{pages}{1--9} (\bibinfo{year}{2025}).
\newblock \urlprefix\url{https://advanced.onlinelibrary.wiley.com/doi/10.1002/admt.202500273}.

\bibitem{Yang2019}
\bibinfo{author}{Yang, L.} \emph{et~al.}
\newblock \bibinfo{title}{{On the Schwarzschild Effect in 3D Two‐Photon Laser Lithography}}.
\newblock \emph{\bibinfo{journal}{Advanced Optical Materials}} \textbf{\bibinfo{volume}{7}}, \bibinfo{pages}{1--9} (\bibinfo{year}{2019}).
\newblock \urlprefix\url{https://onlinelibrary.wiley.com/doi/10.1002/adom.201901040}.

\bibitem{Ocier2020}
\bibinfo{author}{Ocier, C.~R.} \emph{et~al.}
\newblock \bibinfo{title}{{Direct laser writing of volumetric gradient index lenses and waveguides}}.
\newblock \emph{\bibinfo{journal}{Light: Science \& Applications}} \textbf{\bibinfo{volume}{9}}, \bibinfo{pages}{196} (\bibinfo{year}{2020}).
\newblock \urlprefix\url{https://www.nature.com/articles/s41377-020-00431-3}.

\bibitem{Ketchum2022}
\bibinfo{author}{Ketchum, R.~S.}, \bibinfo{author}{Alcaraz, P.~E.} \& \bibinfo{author}{Blanche, P.-A.}
\newblock \bibinfo{title}{{Modified photoresins with tunable refractive index for 3D printed micro-optics}}.
\newblock \emph{\bibinfo{journal}{Optical Materials Express}} \textbf{\bibinfo{volume}{12}}, \bibinfo{pages}{3152} (\bibinfo{year}{2022}).
\newblock \urlprefix\url{https://opg.optica.org/abstract.cfm?URI=ome-12-8-3152}.

\bibitem{Prediger2024}
\bibinfo{author}{Prediger, R.}, \bibinfo{author}{Kluck, S.}, \bibinfo{author}{Hambitzer, L.}, \bibinfo{author}{Sauter, D.} \& \bibinfo{author}{Kotz‐Helmer, F.}
\newblock \bibinfo{title}{High‐resolution structuring of silica‐based nanocomposites for the fabrication of transparent multicomponent glasses with adjustable properties}.
\newblock \emph{\bibinfo{journal}{Advanced Materials}} \textbf{\bibinfo{volume}{36}} (\bibinfo{year}{2024}).
\newblock \urlprefix\url{http://dx.doi.org/10.1002/adma.202407630}.

\end{thebibliography}
\end{document}